\documentclass[aps,prb,twocolumn,superscriptaddress,showpacs,floatfix]{revtex4-1}
\usepackage{epstopdf}
\usepackage{graphicx}
\usepackage{dcolumn}
\usepackage{bm}
\usepackage{bbm}
\usepackage{color}
\usepackage{amsfonts}
\usepackage{amsmath}
\usepackage[normalem]{ulem}
\newcommand\redout{\bgroup\markoverwith{\textcolor{red}{\rule[.5ex]{2pt}{0.4pt}}}\ULon}
\newcommand{\redin}[1]{\textcolor{red}{#1}}

\usepackage[colorlinks=true,citecolor=blue]{hyperref}
\hypersetup{colorlinks=true,citecolor=blue,linkcolor=red,urlcolor=blue}

\begin{document}

\title{Electronic structure and layer-resolved transmission of bilayer graphene nanoribbon in the presence of vertical fields}

\date{\today}

\author{Habib Rostami}
\affiliation{School of Physics, Institute for Research in Fundamental Sciences (IPM), Tehran 19395-5531, Iran}
\author{Reza Asgari}
\email{asgari@ipm.ir} \affiliation{School of Physics, Institute
for Research in Fundamental Sciences (IPM), Tehran 19395-5531,
Iran}
\begin{abstract}
Electronic properties of bilayer graphene are distinct from both
the conventional two dimensional electron gas and monolayer
graphene due to its particular chiral properties and excitation
charge carrier dispersions. We study the effect of strain on the
electronic structure, the edge states and charge transport of
bilayer graphene nanoribbon at zero temperature. We demonstrate a
valley polarized quantum Hall effect in biased bilayer graphene
when the system is subjected to a perpendicular magnetic field. In
this system a topological phase transition from a quantum valley
Hall to a valley polarized quantum Hall phase can occur by tuning
the interplanar strain. Furthermore, we study the layer-resolved
transport properties by calculating the layer polarized quantity
by using the recursive Green's function technique and show that
the resulting layer polarized value confirms the obtained phases.
These predictions can be verified by experiments and our results
demonstrate the possibility for exploiting strained bilayer
graphene in the presence of external fields for electronics and
valleytronics devices.

\end{abstract}

 \maketitle

\section{introduction}

Crystalline bilayer graphene (BLG)~\cite{berger,novo,zhang} has
recently attracted a great deal of attention because of its unique
electronic properties.~\cite{bilayer1} It consists of two
single-layer graphene (SLG) sheets in which the second carbon
layer is rotated by $60^{0}$ with respect to the first one,
separated by a small distance and can be produced by mechanical
exfoliation of thin graphite or by thermal decomposition of
silicon carbide.~\cite{novoselov2006} The low-energy
quasiparticles in BLG behave as massive chiral fermions and are
responsible for a plethora of interesting physics including
broken-symmetry states at very weak magnetic fields when BLG is
suspended to reduce disorder~\cite{vafek} and anomalous exciton
condensation in the quantum Hall regime.~\cite{bilayer2} Although
the intrinsic BLG is a zero-gap semi-metal, it becomes a tunable
band gap semiconductor~\cite{mak, kuzmenko} when a gate voltage is
applied. Much of this attraction is due to its structure having
two Dirac points in the Brillouin zone (BZ), known as $\bm K$ and
$\bm K'$.

Many interesting properties emerge in strained BLG systems in
analogue of monolayer graphene
sheet.~\cite{ref:Ando,ref:Manes,ref:Vozmediano,ref:Guinea,ref:nima,ref:Levy,ref:Rostami}
It has been shown that homogeneous strain profoundly changes the
topology of band structure and the Lifshitz transition takes place
in strained BLG upon splitting the parabolic bands at intermediate
energies into several Dirac cones and consequently this affects
the electron Landau level spectra and the quantum Hall
effect.~\cite{ref:falko11} Moreover, in uniaxially strained BLG,
each ${\bm K}-$point splits into two pockets.~\cite{ref:falko12}
The size of the energy gap can be controlled by adjusting the
strength and direction of different homogeneous strains in
BLG.~\cite{ref:Choi10,ref:Wong12,ref:Raza09} The effect of
interplanar strain on the electronic structure of the bilayer
graphene in the presence of an electric field has been studied by
Nanda and Satpathy~\cite{nanda} within {\it ab inito}
calculations. They found that while strain alone does not produce
a gap, an electric field does so in the Bernel structure but not
in the hexagonal structure. Recently, a theoretical study in
bilayer graphene nanoribbon (BLGNR) shows that a mechanical
deformation can induce effective fields which modifies the
dynamical behavior of electrons.~\cite{ref:Mariani11}
Interestingly, the creation of pseudomagnetic field in a Moir\`{e}
pattern of a twisted BLG has also been reported by Yan and {\it et
al.}~\cite{ref:Yan12} and recently analyzed in STS
measurement.~\cite{stm} A large body of the studies is based on
homogeneously strained BLG even though inhomogeneous strain needs
 specific care.

One of the interesting phenomena that has been observed in bilayer
is the anomalous quantum Hall effect. The quantum Hall effect can
be understood by solving the quantum-mechanical problem for the
bulk of the system in the presence of a magnetic field. There is
an alternative approach to the quantum Hall effect that is
actually based on the analysis of the edge states of electrons in
a magnetic field.~\cite{ref:macdonald84} These edge states are
chiral since only one direction of propagation is allowed. If thus
all bulk states are localized there is still a current being
carried by electrons on the boundary with a contribution to
conductance. BLG is an interesting system to study since its high
magnetic field Landau level consists of equidistant groups of
fourfold degenerate state at finite energy and supports eight
zero-energy states and can be broken by perpendicular
electric field.~\cite{bilayer2, review} This helps to control the
edge state structure using a combination of the electric field and
the magnetic field too.

In the BLG, there are four carbon atoms per unit cell and it turns
out that there is also (at least for the case that energies are
below the interlayer bonding energy $\gamma_1$) a spin $1/2$
pseudospin degree of freedom can be interpreted as labeling a
layer rather than a sublattice. Since the sublattice pseudospin is
equivalent to a layer, an electric field perpendicular to layers
couples to the pseudospin degree of freedom much stronger than a
practical magnetic field coupled to the real spin. This surprising
feature of the electronic properties of BLG offers a possible
potential for digital electronics based on graphene
flakes.~\cite{ref:MacDoanld12,ref:Jose09,ref:Li10} Moreover, in
the BLG system there is a length scale corresponding to the interlayer
hopping that indicates the traveling length of electrons between the
interlayer hopping and it is about $l_{\perp}\approx11a_0$ with
$a_0=0.142$nm. Therefore, a layer-flip relaxation time $\tau_{Lf}$
should be defined in analogue with a spin-flip relaxation time in
non-collinear spin dependence transport. Typical length of bilayer
device is much longer than the $l_\perp$, in order that two layers
are strongly coupled.~\cite{ref:Beenakker07}

The general features of the electronic structure, basically the
structure of the Landau level and edge states in the quantum Hall
regime, have been numerically studied~\cite{ref:fertig} for the
unbiased and unstrained BLGNR. In the present work we study the
effect of strain on the electronic structure, particulary the
edge states and charge transport, of the BLGNR system at zero
temperature. In the system, similar to monolayer graphene, a
pseudomagnetic field can be created. We first propose a lattice
model Hamiltonian to explore BLG under the combined effect of
deformations, real magnetic fields and gate voltages. The lattice
model Hamiltonian, which we propose, provides a quite good
description of the edge states since it incorporates the correct
expressions of the hopping integrals when the system is deformed
and we show that pseudo Landau levels in the bulk are no longer
dispersive. We study comprehensively the edge states inside the
gap (created by vertical bias) in the presence of a real and
pseudomagnetic fields and explore different phases in BLGNR for
the different values of the interlayer spacing. We obtain a valley
polarized quantum Hall effect in biased bilayer graphene when the
system is subjected to a perpendicular real magnetic field for
certain values of the interlayer hopping integral. We show that,
in stained and biased BLGNR exposed by a magnetic field, a
topological phase transition from a quantum valley Hall to a
valley polarized quantum Hall can occur by tuning the interlayer
spacing (interplanar strain) between the two layers. In different
phases, the number of conducting channel is calculated by
recursive Green's function approach together with the Landauer
formalism and the conductance feature supports the existence of
those phases. Moreover, we investigate the effect of the
interlayer spacing on the layer-resolved transport in BLGNR by
using an analog between layer and spin-resolved transport
physics.~\cite{ref:Pareek02}

The paper is organized as follows. In Sec.~II we introduce the
formalism that will be used in calculating the electronic
structure, two terminal conductance and the layer polarization
quantity from the recursive Green's function approach. In Sec.~III
we present our analytic and numeric results for the dispersion
relation in both strained and unstrained biased bilayer graphene
sheets in the presence of the magnetic field. Section~IV contains
a brief summary of our main results.

\section{Theory and Model}\label{sec:theory}

\begin{figure}
\includegraphics[width=0.7\linewidth]{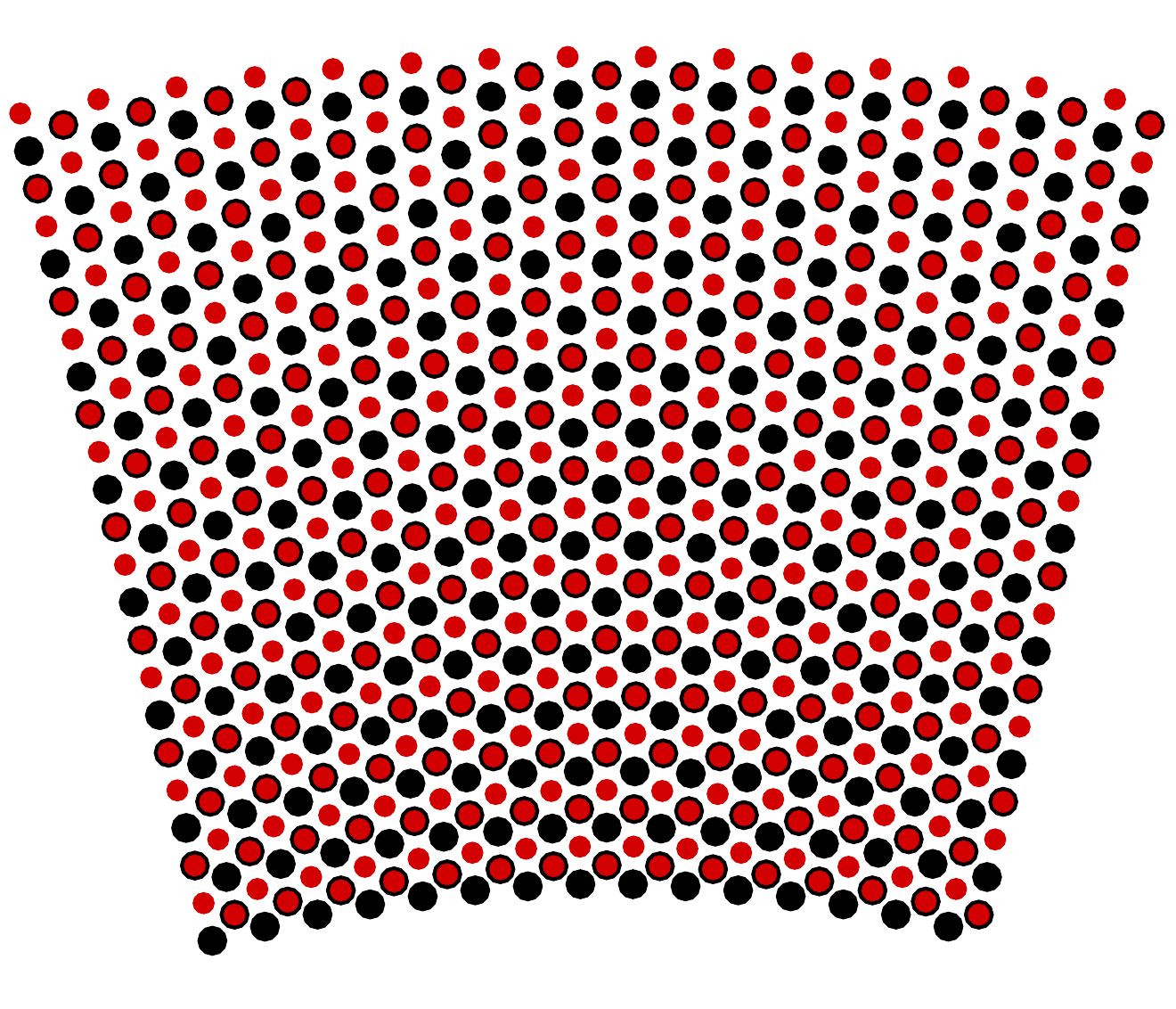}
\includegraphics[width=0.25\linewidth]{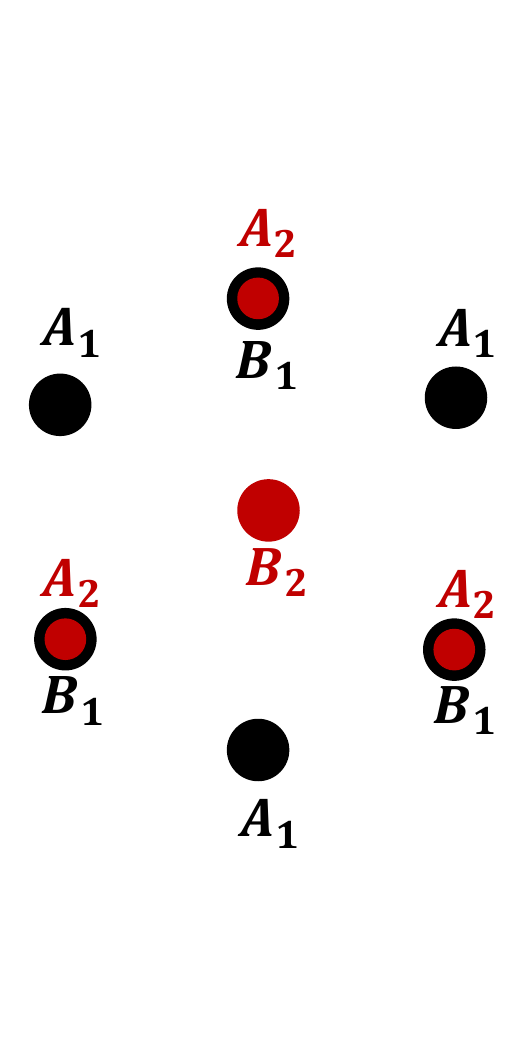}
\caption{(Color online) Sketch of an arc-shaped bending bilayer graphene lattice and sublattice labels
in a Bernal-stacking and the top view of the positions of carbon atoms.}
\label{fig1}
\end{figure}

Bilayer graphene is a crystal system consisting of two single layers
of graphene sheet offset from each other in the $xy$ plane. The
top $A$ sublattice is directly above the bottom $B$ sublattice and
it is between these pairs of atoms that the interlayer dimer bonds
are formed, whereas there are no essential hopping processes
between a counterpart on the other layer. We consider an
arc-shaped bending BLGNR with zigzag edges to explore the
electronic structures in the presence of perpendicular fields. We
assume that the $sp^2$-hybridized electrons of carbon atoms in
each sheet are inert and only take into account the $2p_z$
electrons which form the $\pi$ bands. A top view of the system is
displayed in Fig.~\ref{fig1} and hopping parameters are defined as
$t_{A_1,B_1}=t_{A_2,B_2}=\gamma_0$, $t_{B_1,A_2}=\gamma_1$,
$t_{A_1,B_2}=\gamma_3$ and $t_{A_1,A_2}=\gamma_4$.

Most of the interesting properties of the Bernel-stacking BLGNR
can be understood in the framework of a tight-binding model
Hamiltonian.~\cite{bilayer2} In deformed graphene system, the
ideal atomic configurations change and all bonds of each atom with
its neighbors are no longer equivalent and therefore the hopping
parameters are different throughout the whole sample. We modify
the tight-binding model for a deformed BLGNR by taking into
account atomistic inhomogeneities, and we find the following
effective Hamiltonian
\begin{eqnarray}\label{eq:real-H}
H&=&\sum_{i,l}\{{\epsilon_i}^i{a_i^l}^\dagger
a_i^l+{\epsilon_i}^b {b_i^l}^\dagger b_i^l\}\\ \nonumber
&-&\sum_{<ij>,l}\{\gamma^{ij}_0e^{\phi_{ij}}{a_i^l}^\dagger
b_j^l+H.c.\}\\ \nonumber &-&\sum_{i}\{\gamma^{ij}_1{a_i^2}^\dagger
b_i^1+H.c.\}\\ \nonumber
&-&\sum_{<ij>}\{\gamma^{ij}_3e^{\phi_{ij}}{a_i^1}^\dagger b_j^2+
H.c.\}
\end{eqnarray}
where we define~\cite{ref:Moon12,ref:Magaud10,ref:ZFWang12}
\begin{eqnarray}
\gamma^{ij}_0&=&\gamma_0e^{-\beta_0(\frac{d}{a_0}-1)},~~~~~~~~~\gamma^{ij}_1=\gamma_1e^{-\beta_1(\frac{d}{c_0}-1)}\nonumber\\
\gamma^{ij}_3&=&X\left[\gamma_0(\frac{d_{\parallel}}{d})^2e^{-\beta_0(\frac{d}{\tilde{c}_0}-1)}+
\gamma_1(\frac{d_{\perp}}{d})^2e^{-\beta_1(\frac{d}{\tilde{c}_0}-1)}\right]\nonumber\\
\end{eqnarray}
with $\gamma_0=-3.16$eV, $\gamma_1=0.39$eV, $a_0=0.142$nm,
$c_0=0.335$nm, $\tilde{c}_0=\sqrt{a^2_0+c^2_0}=0.364$nm,
$\beta_0=2.9$, $\beta_1=c_0\beta_0/a_0=6.8$ and $d=|{\bf r}_i-{\bf
r}_j|$ which has in-plane($d_{\parallel}$) and
vertical($d_{\perp}$) components. $\gamma^{ij}_3$ in Eq.~(2) results from the Slater-Koster table~\cite{slater54} with an
empirical parameter. Setting $X=1.76$, it generates a reasonable
hopping energy for an ideal BLG with $\gamma_3=0.315$eV. In
addition, ${\epsilon_i^{a(b)}}$ is the on site potential at point
$A$($B$). In a clean ribbon, we have $\epsilon_i=(-1)^lu/2$ where
$u$ stands for induced potential due to external perpendicular
electric field which breaks layer degeneracy and $l=1,2$ stands to
the layer number.
$\phi_{ij}=\frac{e}{\hbar}\int_i^j{\vec{A}.\vec{dr}}$ is the
Peierls phase factor to carry out the orbital effect of the
perpendicular real magnetic field. Another hopping energy between
the nearest-neighboring layers, $\gamma_{4} = 0.04 eV$, is very
small compare to $\gamma_{0}$ and can be ignored.

It would be worthwhile mentioning that the exponential correction
of the hopping integrals ($\beta_0$ term) is consistent with
experimental results~\cite{ref:exp1} of SLG in which
$\partial\gamma_0 /\partial a_0=-6.4$ eV{\AA}$^{-1}$ and it
provides quite good descriptions of the system with respect to
those results obtained within Harrison's
approach~\cite{ref:Pereira} in which
$\gamma^{ij}_{0(1)}=\gamma_{0(1)}[1-\beta_{0(1)}{\bm
d}_0\cdot({\bm d}-{\bm d_0})/d_0]$ where $\beta_{0(1)}\approx 2$
and ${\bm d}_0$ is the distance between two atoms in unstrained
graphene. We will discuss the discrepancy of these approaches in
the next Section and show that pseudo Landau levels in the bulk
are no longer dispersive by using the exponential correction of
the hopping integrals.

In the absence of strain and external magnetic field the ${\bm
k}$-space Hamiltonian in four dimensional sublattice space of
$(A_1,B_2,A_2,B_1)$ can be written as follows
\begin{eqnarray}
{\cal H}=\begin{pmatrix}{\cal H}_1&{\cal V}\\{\cal V}^\dagger&{\cal H}_2\end{pmatrix}
\end{eqnarray}
where the interlayer and intralayer contributions are ${\cal
H}_{i=1,2}=U_{i}\sigma_0+\hbar v_0(\sigma_x q_x+\sigma_y q_y)$ and
$2{\cal V}=\gamma_1 \sigma_{-}+v_3(q_x+i q_y)\sigma_{+}$,
respectively. Here $\sigma_{\pm}=\sigma_x i\pm \sigma_y$ with
Pauli matrix $\sigma_i$. This Hamiltonian includes a non-collinear
physics for the layer degree of freedom which it can be assumed as
an pseudospin in BLG. ~\cite{ref:MacDoanld12} A general
Hamiltonian for the deformed BLG Hamiltonian in the ${\bm
k}$-space is quite cumbersome~\cite{ref:Mariani11} however, in the
low-energy limit, it can be simplified for some special cases. For
instance, when the deformation is just in-plane where two layers
are deformed in a same way, the two-band Hamiltonian can be
written as
\begin{eqnarray}\label{eq:p-H}
H_k&=&-\frac{v_0^2}{\gamma_1}
\begin{pmatrix}
0&({\bf p}-e{\bf A})^{\dagger^2}\\({\bf p}-e{\bf A})^2&0\end{pmatrix}\nonumber\\\nonumber\\&+&\begin{pmatrix}
\frac{u}{2}&v_3({\bf p}-e\eta_3{\bf A})\\v_3({\bf p}-e\eta_3{\bf A})^\dagger&-\frac{u}{2}.
\end{pmatrix}
\end{eqnarray}

Here $v_0=3\gamma_0 a_0/2\hbar$, $v_3=3\gamma_3 a_0/2\hbar$,
$\eta_{3}=2({a_0}/{\tilde{c}_0})^2{\beta_{3}}/{\beta_0}$,
$\beta_0=-{\partial \log \gamma_0}/{\partial \log a_0}$ and
$\beta_{3}=-{\partial \log \gamma_{3}}/{\partial \log
\tilde{c}_0}$. The fictitious gauge filed, therefore, is defined
as ${\bf
A}=\frac{\phi_0}{a_0}\frac{\beta_0}{2\pi}(\varepsilon_{xx}-\varepsilon_{yy},-2
\varepsilon_{xy})$. Notice that $\eta_{3}\approx2$ based on our
model. In this paper, however, we would like to consider more
general cases and thus we use Eq.~(\ref{eq:real-H}) in the real
space Hamiltonian and find its associated Hamiltonian in the ${\bm
k}-$space. In ribbon geometry, in other words, it is easy to find
the energy dispersion relation with the periodic boundary
condition along the ribbon in the $x-$direction. To do so, we assume
an infinite stack of principal layers with the nearest-neighbor
interactions. A principle layer is defined as the smallest group
of neighboring atoms planes in such a way that only nearest-neighbor
interactions exist between principle layers. Thus, we can
transform the original system into a linear chain of the principal
layers.~\cite{ref:Hatami11} Owing to the translational invariant
along $x$, the momentum in the $x-$direction is a good quantum
number. To study the band structure properties provided by our
tight-binding model, we find its ${\bm k}-$space forms as
$\sum_{{\bm k}}\psi_{{\bm k}}^{\dag} H_k \psi_{{\bm k}}$. After
performing the Fourier transformation along the $x-$direction, the
Hamiltonian in ${\bm k}-$space can be written as
\begin{equation}\label{eq:H-k1}
 H_k=H_{00}+H_{01}e^{-i k_x
a}+H_{01}^\dagger e^{i k_x a}
\end{equation}
in which $a=\sqrt{3}a_0$. Moreover, $H_{00}$ and $H_{01}$ describe
coupling within the principal layer (intra layer) and the adjacent
principal layers (inter layer), respectively based on the
tight-binding model given by Eq.~(\ref{eq:real-H}). This
Hamiltonian can be simply diagonalized and thus the energy
dispersion of a ribbon in the presence of a real and
pseudomagnetic fields can be obtained. Furthermore, corresponding
wave function for a given energy and wave vector can be used in
order to calculate the site-resolved local density of states
(LDOS) for the ribbon as $\rho(y,E_{nk})=\sum_{m k'}{|\psi_{m
k'}(y)|^2 \delta(E_{n k}-E_{m k'})}$ where $n,m$ are band indexes.

In general, strain can induce a scalar potential as a diagonal
term~\cite{ref:Mariani11} and it originates from a redistribution
of electron density under the deformation. Since the scalar
potential suppresses drastically due to the effect of electron
screening~\cite{ref:Katsnelson12} and describes only the electron
profile on the BLG, we thus ignore the scalar potential in our
calculations. We assume an arc-shape deformation in order to
produce fictitious gauge field in which the atomic displacement
profile is $(u_x,u_y)=({xy}/{R},-{x^2}/{2R})$ in both layers. This
deformation leads to a constant pseudomagnetic field with opposite
sign in two valleys as $|B^{ps}|={\phi_0 \beta_0}/{2\pi a_0 R}$ (
by neglecting $\gamma_3$) with corresponding magnetic length
$l_B^{ps}=\sqrt{{2 a_0 R}/{\beta_0}}$. Here, $R$ is the bending
radius of the deformation applied to the flakes.

Importantly, the fictitious gauge field appearing in the
arc-shaped in BLGNR differs from that created in the arc-shaped
bending SLG nanoribbon due to the interlayer hopping
contributions. The concept of the pseudomagnetic field is still
well-defined in BLGNR as well since the interlayer contribution of
the fictitious gauge field is weaker than its counterpart for the
intralayer hopping. Moreover, the variations of $\gamma_1$ is
controllable by a perpendicular force which changes the interlayer
spacing without generating any pseudomagnetic field. In the vertical
direction, the mechanical strength generates from a weak
$\sigma$-bond between $p_z$ orbitals of carbon atoms, we can thus
change the interlayer spacing in a wide range without breaking the
symmetries of the system. We reveal the effect of the interlayer
hopping values by looking at the physics of the edge states in the
system.

To calculate conductance we use the non-equilibrium Green's function
method in which retarded Green's function is defined as
$G=(E-H-\Sigma+i0^+)^{-1}$ by employing the recursive Green's
function method.~\cite{ref:negf02} In noninteracting Hamiltonian
the self-energy ($\Sigma=\Sigma_L+\Sigma_R$) only originates from
the connection of the system to the leads and it can be calculated
by a method that has been developed and implemented for
BLGNR.~\cite{ref:lopez, ref:Hatami11} Transmission function, $T$
is obtained from Green's function and line width function
$\Gamma_{L,R}$ too.~\cite{ref:Hatami11} The Landauer formula then
gives us the zero temperature conductance, $2e^2 T/h$. In order to
study the layer-resolved transport of BLGNR, we have designed an
electronic system in which leads are modelled as two decoupled
monolayer graphene. In this case $\Gamma_{L,R}$ are block diagonal
matrixes in layer-space and therefore layer-resolved transmission
components at zero temperature can be read as follows
\begin{eqnarray}\label{eq:T}
T^{\sigma\sigma'}&=&{\rm Tr}[\Gamma^{\sigma\sigma}_L G^{\sigma\sigma'} \Gamma^{\sigma'\sigma'}_R {G^\dagger}^{\sigma'\sigma}]\\ \nonumber
\Gamma_{L,R}&=&-2\Im m[\Sigma_{L,R}]
\end{eqnarray}
where $\sigma,\sigma^{'}$ correspond to the layer degree of
freedom and being $1$ or $2$. In Landaure formalism,
zero-temperature conductance in a two terminal setup is given by
\begin{equation}\label{eq:G}
G=\frac{2e^2}{h}(T_{Lc}+T_{Lf})
\end{equation}
where layer-conserve and layer-flip transmissions are defined as
$T_{Lc}=T^{11}+T^{22}$ and $T_{Lf}=T^{12}+T^{21}$, respectively.
These definitions are given in a similar way that the
spin-resolved components have been specified.~\cite{ref:Pareek02}
We thus define the layer polarization, in analog with
non-collinear spin-dependent system, as
$P=(T_{Lc}-T_{Lf})/(T_{Lc}+T_{Lf})$ which can be valued between
$1$ and $-1$ corresponding to a completely layer-conserve and
layer-flip transport cases, respectively. In order to count the
number of transport channel\redin{s}, we assume the same structure of the
leads and scattering region in order that unexpected scattering
can be eliminated in the interface of the leads and scattering
region.

\begin{figure}
\includegraphics[width=0.49\linewidth]{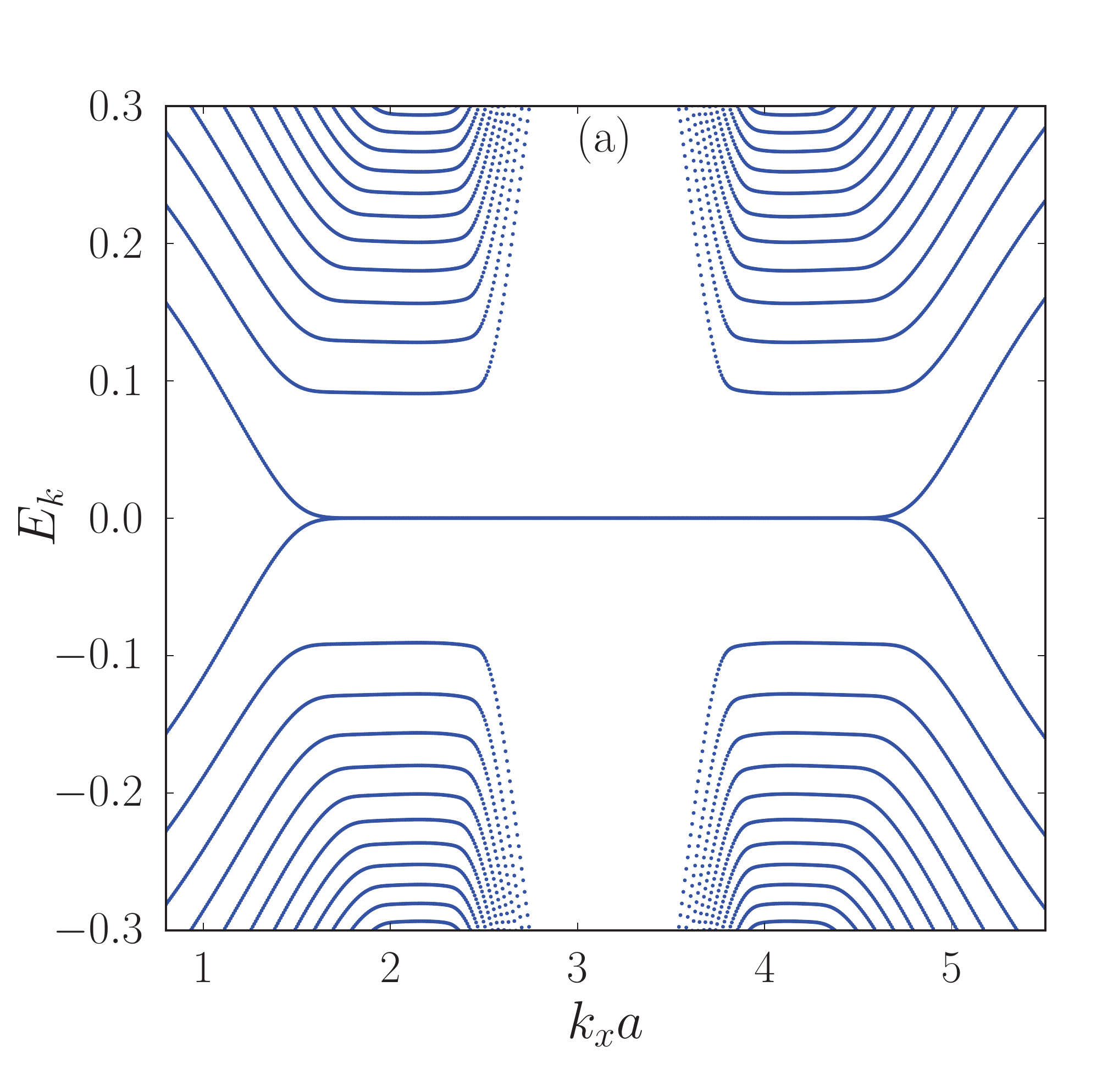}
\includegraphics[width=0.49\linewidth]{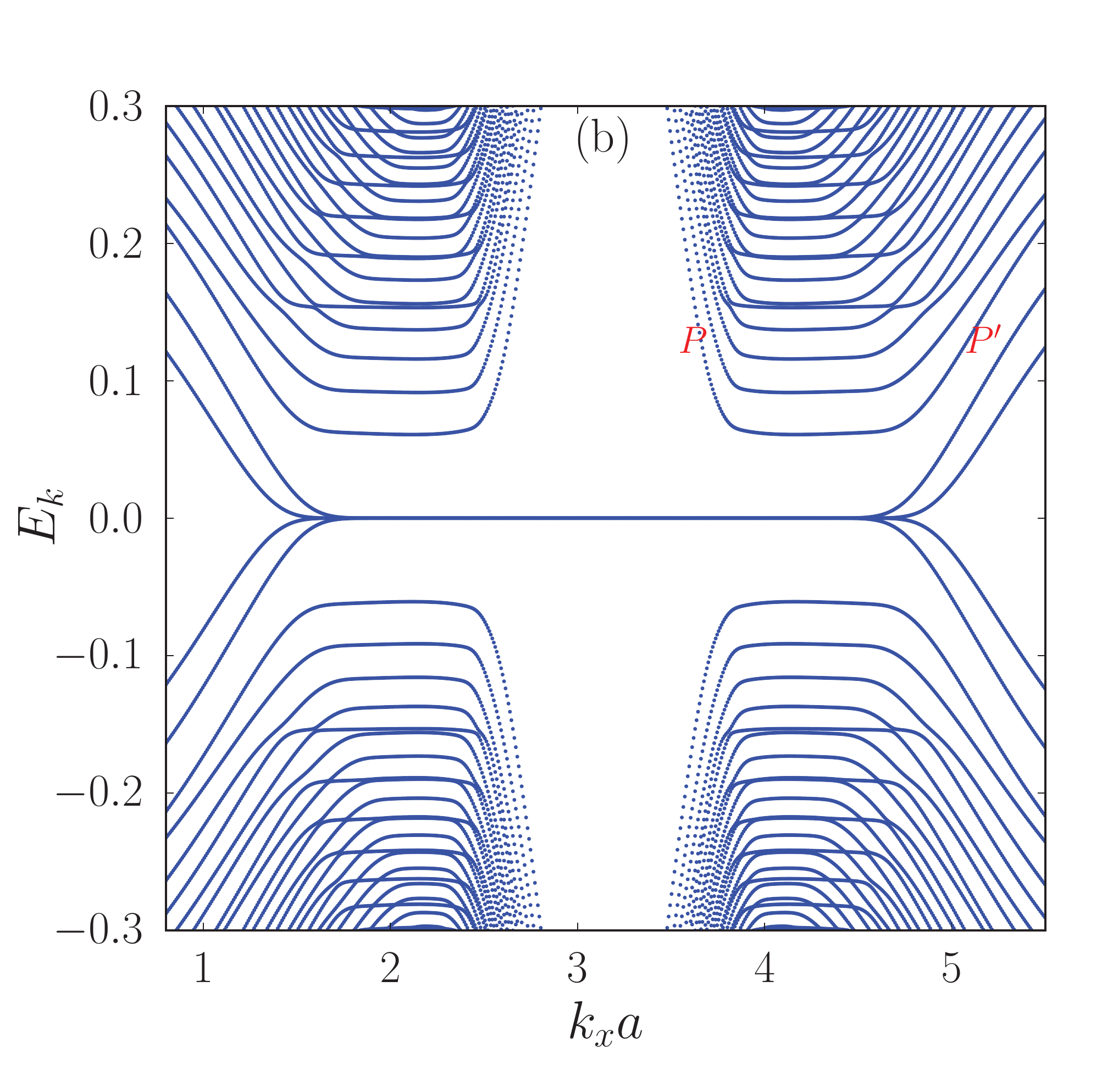}
\caption{(Color online) Pseudo Landau level in a SLG nanoribbon (a) and BLGNR (b) as a
function of $k_x a$ for the ribbon width $W=300.5 a_0$ and the bending radius $R=601a_0$. The PLLs are no longer
dispersive since the exponential hopping correction used in our model results in flat
PLLs. The zigzag boundary condition do
not mix the Dirac points leading to a complete distinction of the
edge modes. Two edge states on two edges of the arc-shaped graphene are no longer the same (
labelled by $p$ and $p'$ in (b)) thus
quasiparticles in one edge move faster than one along another
edge.}
\label{fig2}
\end{figure}
\begin{figure}
\includegraphics[width=0.49\linewidth]{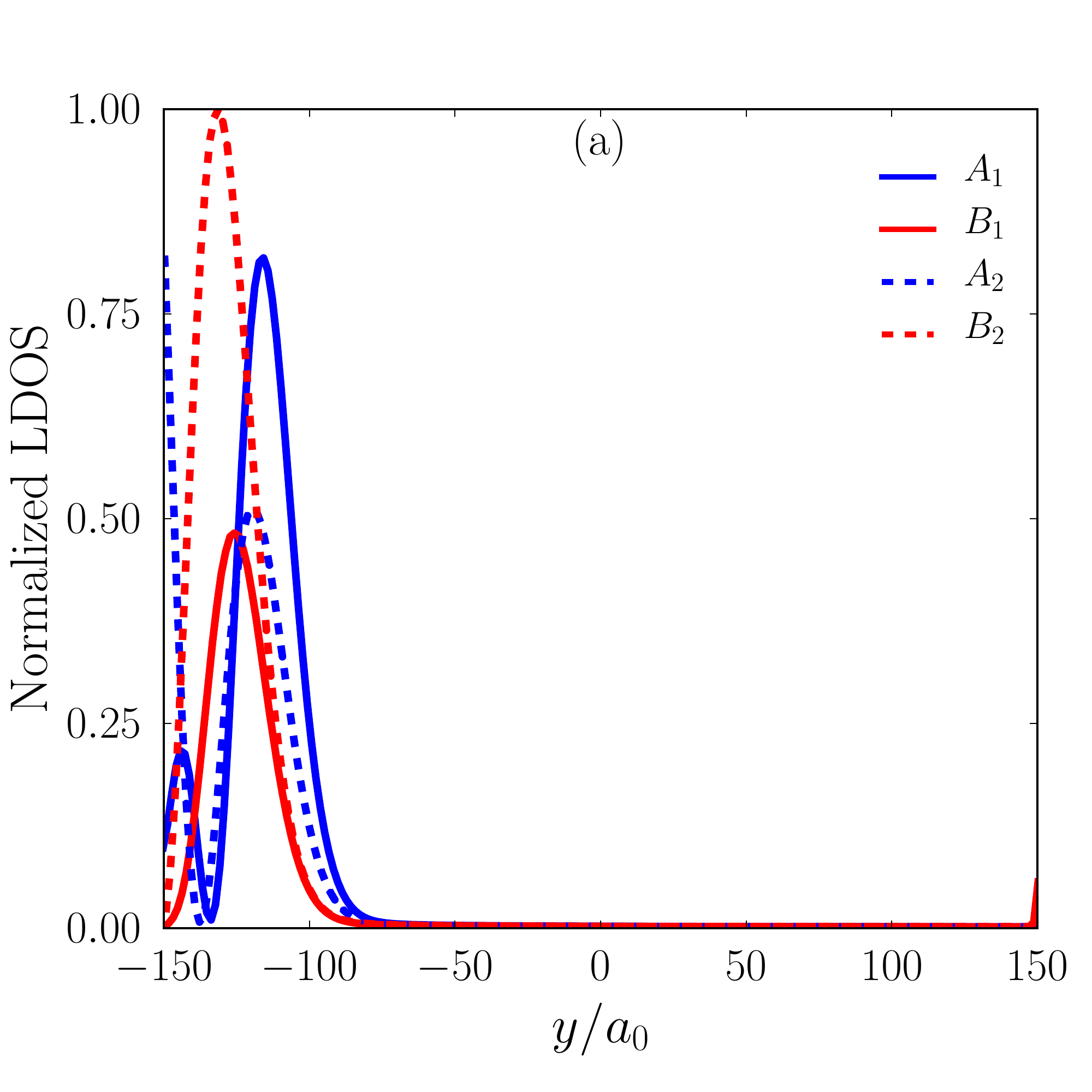}
\includegraphics[width=0.49\linewidth]{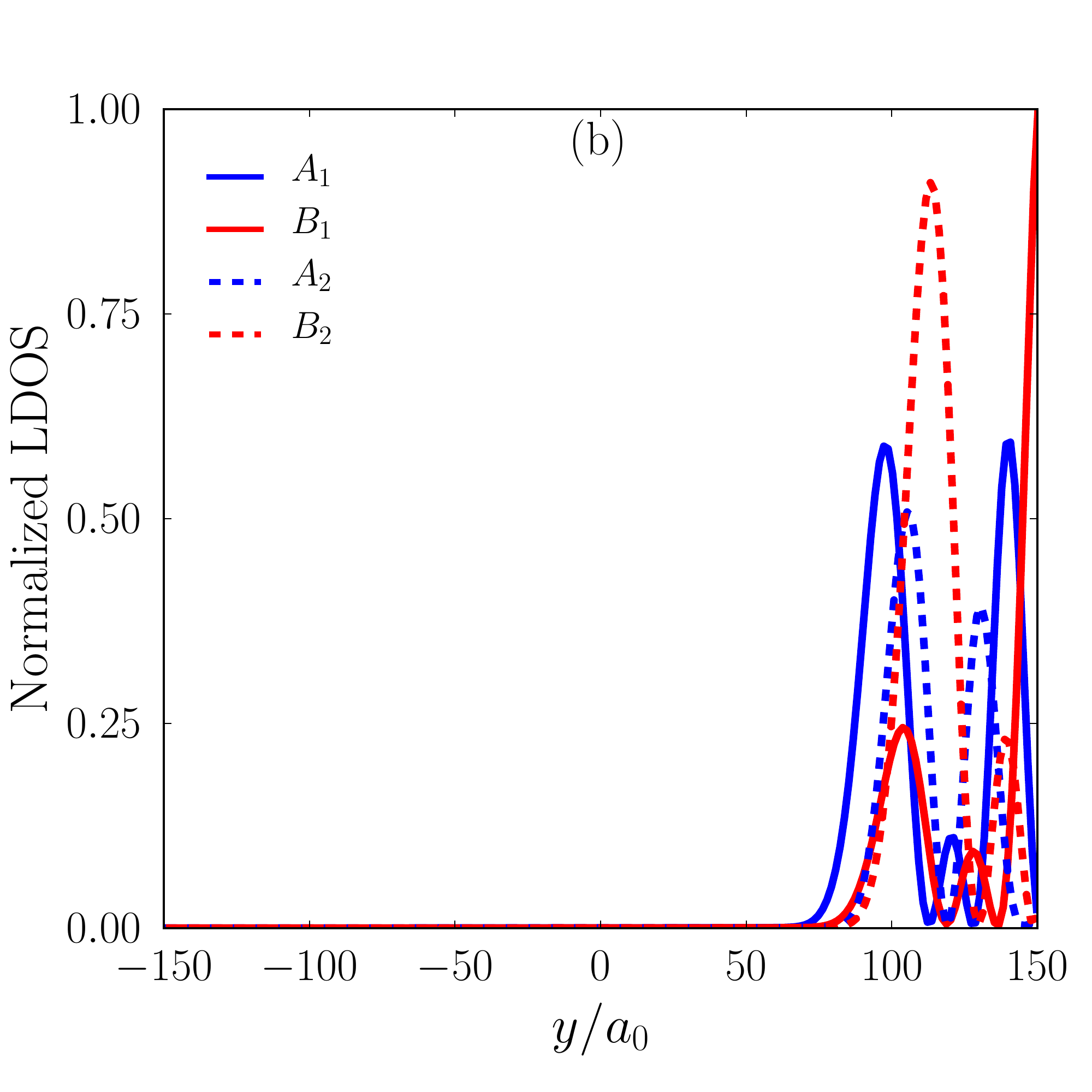}
\caption{(Color online) Site-resolved normalized local density of states (LDOS) as a function
of distance for PLLs for P (a) and $P^\prime$ (b) edge state indicated in Fig.~2 b
for $E=0.13t_0$, $k_P=3.7$ and $k_{p'}=5.1$. Numerical results
show that the faster moving particle on edge is localized
on the compressed edge and the slower one is located on the stretched
edge.}
\label{fig3}
\end{figure}

\section{Numerical results and Discussions}

In this section, we present our main calculations for the
electronic properties of BLGNR by evaluating
Eqs.~(\ref{eq:real-H}), (\ref{eq:H-k1}) and (\ref{eq:T}). The
general features of the electronic structure, basically the
structure of the Landau level and edge states in the quantum Hall
regime, have been numerically studied previously~\cite{ref:fertig}
for the unbiased and unstrained BLGNR. We first modify the lattice
Hamiltonian by using correct expressions of the hopping integrals
for strained graphene. Second, we present our comprehensive
numerical results of electronic structure by exploring the edge
states and the layer-resolved transport properties of biased BLGNR
in two different interesting regimes an ideal and a
deformed bilayer graphene.

To investigate the generation of pseudomagnetic field in the
arc-shaped bending SLG and BLG nanoribbon systems we first
calculate the electronic spectrum shown in Fig.~\ref{fig2} in
which the flat pseudo Landau levels (PLLs) can be clearly
obtained. It should be noticed that PLLs in the bulk are not
dispersive and our numerical results differ slightly from
dispersive PLLs results reported in
Refs.~[\onlinecite{ref:Low10},\onlinecite{ref:Roy12}]. Dispersive
Landau levels are just expected in the system when it is subjected
by an electric field which generates a drift velocity of charge
carriers through the edge of the system.~\cite{Lukose07} The
reason for the discrepancy mainly arises from the different form
of the hopping corrections. Based on our numerical calculation,
the Harrison's method results in dispersive
PLLs~\cite{ref:Low10,ref:Roy12} however the exponential hopping
correction that we have implemented in our model results in 
flat PLLs. The spectrum in both SLG and BLG nanorribbon systems
is not equidistance. The zigzag boundary condition does not mix the
Dirac points leading to a complete distinction of the edge modes.
Furthermore, the zero Landau level with zero-energy is chiral and
topologically protected. Meanwhile, two edge states on the two
edges of the arc-shaped graphene are no longer the same (labeled
by $p$ and $p'$ in Fig.~\ref{fig2}b) so that quasiparticles in one
edge move faster than one along other edge. This difference arises
from compressing and stretching strain in the bottom and top edges
respectively and as a result the velocity of quasiparticle in the
compressed edge is larger than the one in the stretched
edge.~\cite{ref:Rostami} In order to clarify this analysis, we
thus show in Fig.~\ref{fig3} the site-resolved local density of
state of those edge states and our numerical results confirm that
the faster moving particle on edges is localized on the compressed
edge and the slower one is located on the stretched edge.

\begin{figure}
\includegraphics[width=1\linewidth]{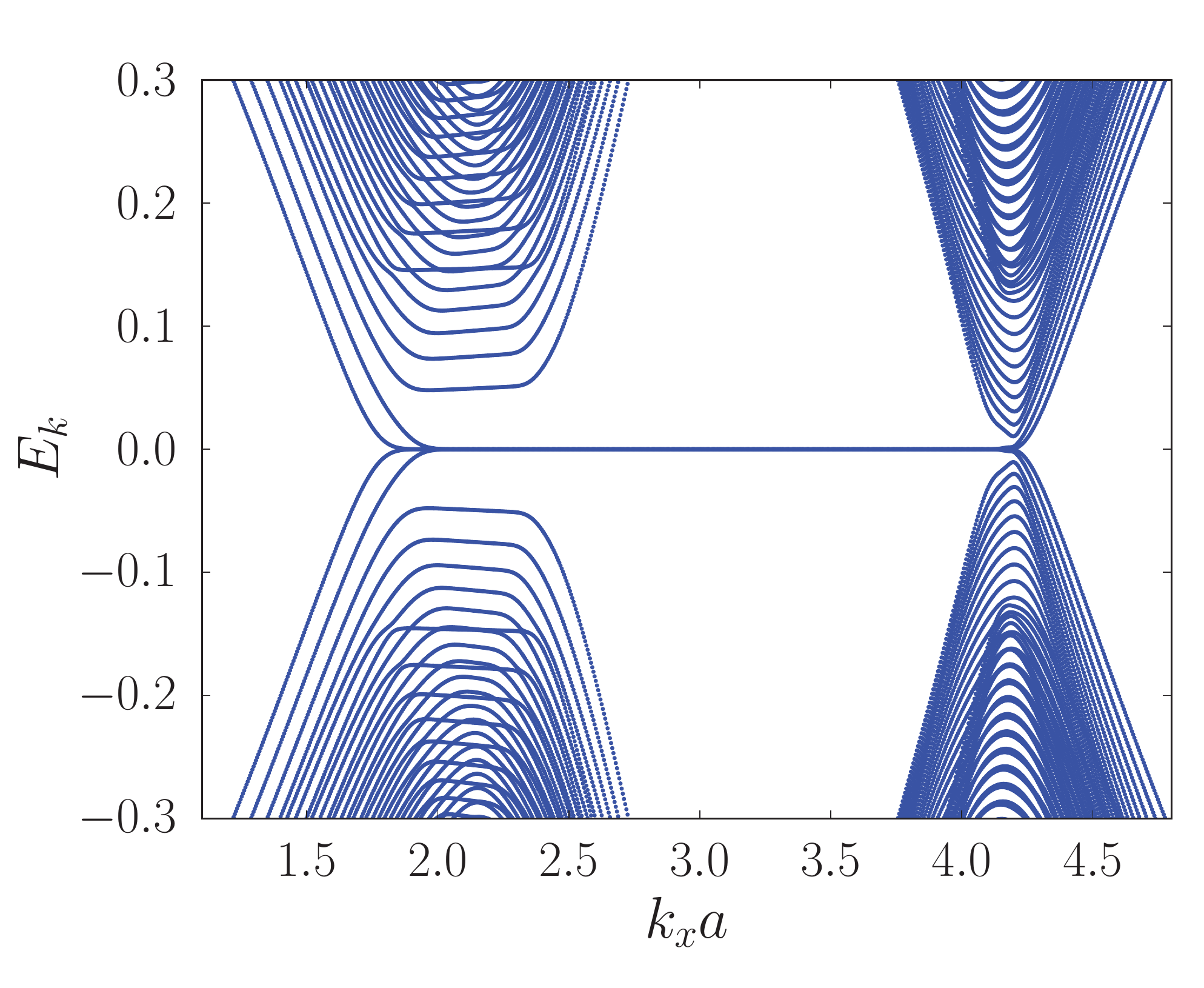}
\caption{(Color online)
Energy dispersion in an arc-shaped bending BLGNR in a real magnetic field as a function of $k_x a$
for $W=300.5a_0$, the bending radius $R=1803a_0$ and
the pseudomagnetic length $l_B=40.5a_0$. Clear quantum Hall plateaus are
observed mimicking the conventional quantum
Hall effect in BLG. The combination of the pseudomagnetic field
due to deforming and a real magnetic field leads to the
effective total fields acting on electrons from the valleys ${\bm
K}$ and ${\bm K'}$ being different which results in a valley
polarization.}
\label{fig4}
\end{figure}
\begin{figure}
\includegraphics[width=1\linewidth]{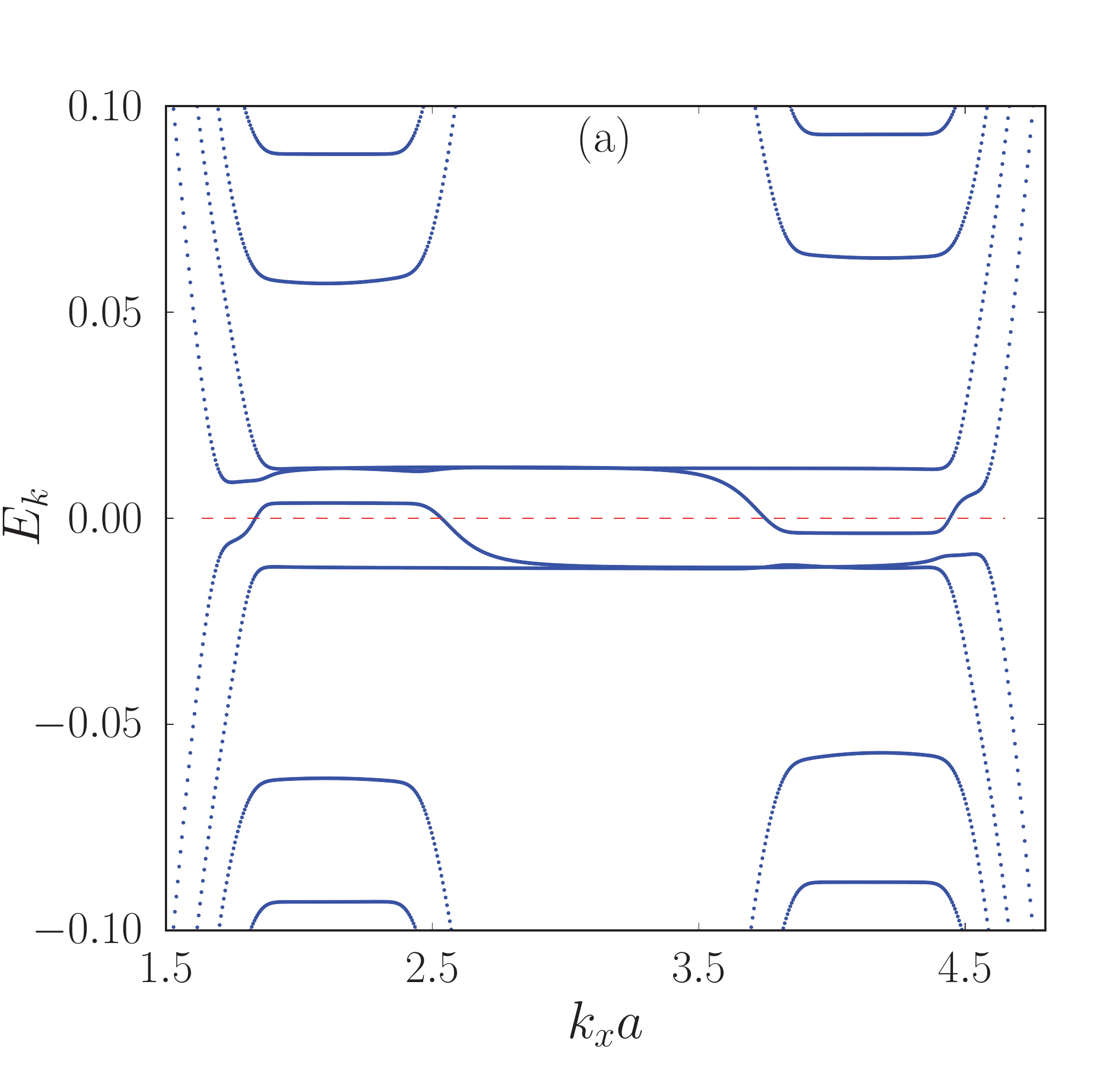}
\includegraphics[width=1\linewidth]{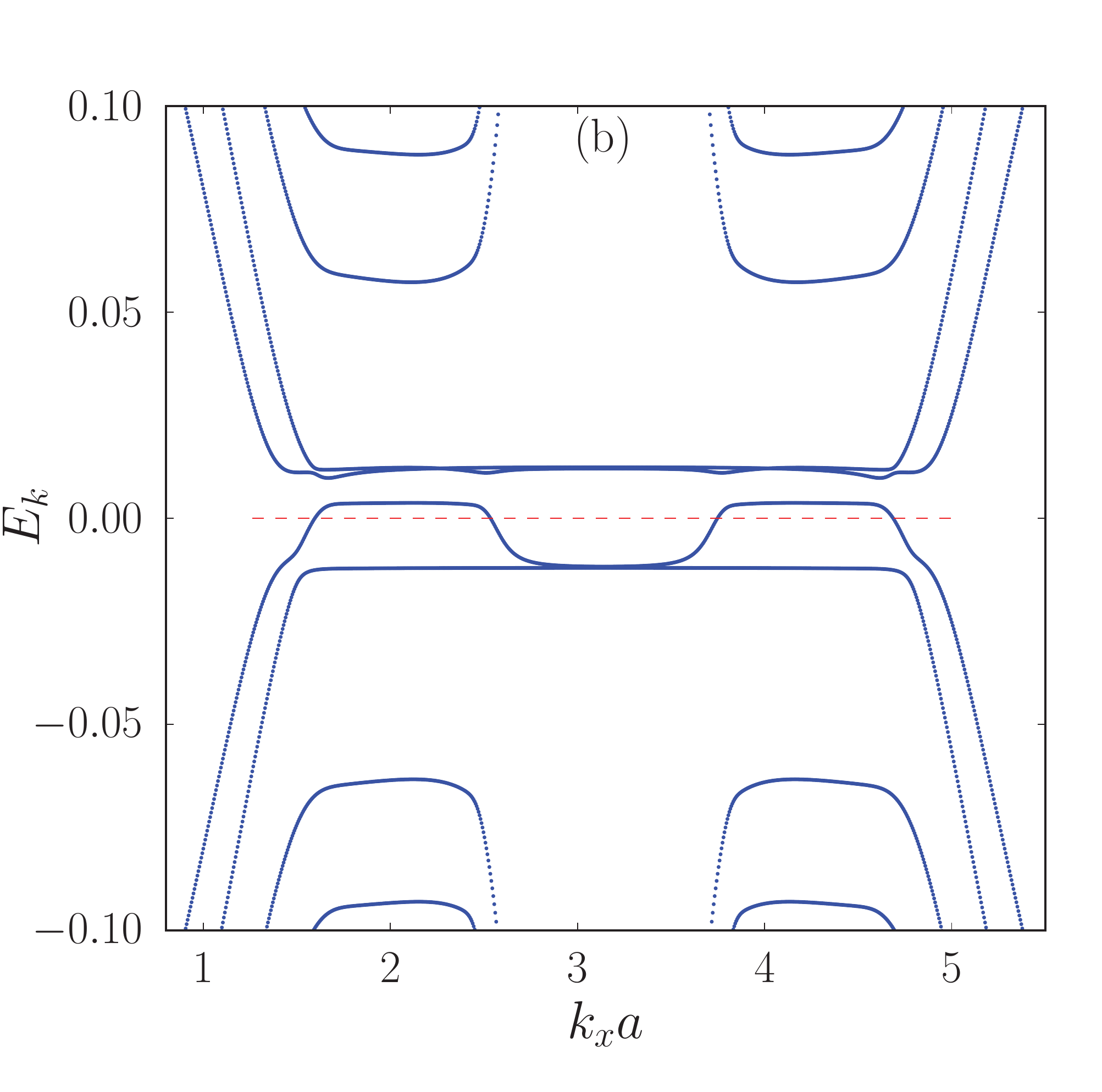}
\caption{( Color online) Landau levels in a biased BLGNR (a)
and pseudo Landau levels in an arc-shaped bending of the biased BLGNR (b)
for $W=300.5a_0$, $R=601a_0$, $l_B=23.4a_0$, the bias gate voltage $u=\gamma_1/5$.
Dashed red line indicates a constant energy level. Due to the breaking of layer
degeneracy by applying bias, the zero Landau level splits into two
levels and then higher level shifts in the opposite direction near 
two valleys when the BLGNR is subjected by the real magnetic field.
For the case of pseudo
Landau levels there is a mirror
symmetry for the energy dispersion of the two valleys and the
level shift of pseudo Landau levels is in the same
direction in the valleys.}
\label{fig5}
\end{figure}

Next, we examine the interplay between real and pseudomagnetic
fields. We consider strained BLGNR by deforming it in the
arc-shaped bending structure. We calculate the energy spectrum of
the system in the presence of a real magnetic field and the
electron spectrum is shown in Fig.~\ref{fig4}. Decent quantum Hall
plateaus are observed mimicking the conventional quantum Hall
effect in BLG. The combination of the pseudomagnetic field due to
the deformation and a real magnetic field leads to a broadening of
all Landau levels except the zero-energy. Moreover, the effective
total fields acting on electrons from the valleys ${\bm K}$ and
${\bm K'}$ are different which results in a valley polarization.
As we mentioned before, the pseudomagnetic field picture in BLG is
no longer the same as in SLG, because in strained SLG the momentum
${\bf p}$ tends to ${\bf p}+e{\bf A}$, where ${\bf A}$ is the
fictitious gauge field. In BLG, however, the momentum on
interlayer hopping feels like a different pseudo gauge field. Also the
energy of PLLs decreases with increasing interlayer vertical
hopping consistent with the energy of LLs in BLG which is given by
$E_n=\frac{2}{\gamma_1}(\frac{\hbar v_{\rm
F}}{l_B})^2\sqrt{n(n-1)}$.

It is worthwhile studying the PLLs and LLs in the presence of a
perpendicular gate voltage. We calculate the energy dispersions of
biased BLGNR in the presence of a real and pseudomagnetic fields
and the results are illustrated in Figs.~\ref{fig5}a and
Fig.~\ref{fig5}b, respectively. Due to the breaking of the layer
degeneracy by applying the bias gate, the edge state at zero
Landau level splits into two levels and then higher level shifts
in the opposite direction near to each valley. For the case of PLLs in
biased BLGNR, however different features can be occurred which
originates from the time reversal symmetry in the mechanically
deformed lattice. In this case, there is a mirror symmetry for the
energy dispersion of the two valleys and the level shift of PLLs
is in the same direction in the valleys. This shift is equal to
$\frac{\hbar e u}{2\gamma_1}\xi B_\xi$ in which $\xi=\pm$ denotes
the valley degree of freedom and $B_\xi=\xi |B_{ps}|$ and thus the
PLLs shift equally downward for $u>0$ near to the position of the
valleys. The calculated results shown in Fig.~\ref{fig5} are in
good agreement with these analysis.

\begin{figure}
\includegraphics[width=0.49\linewidth]{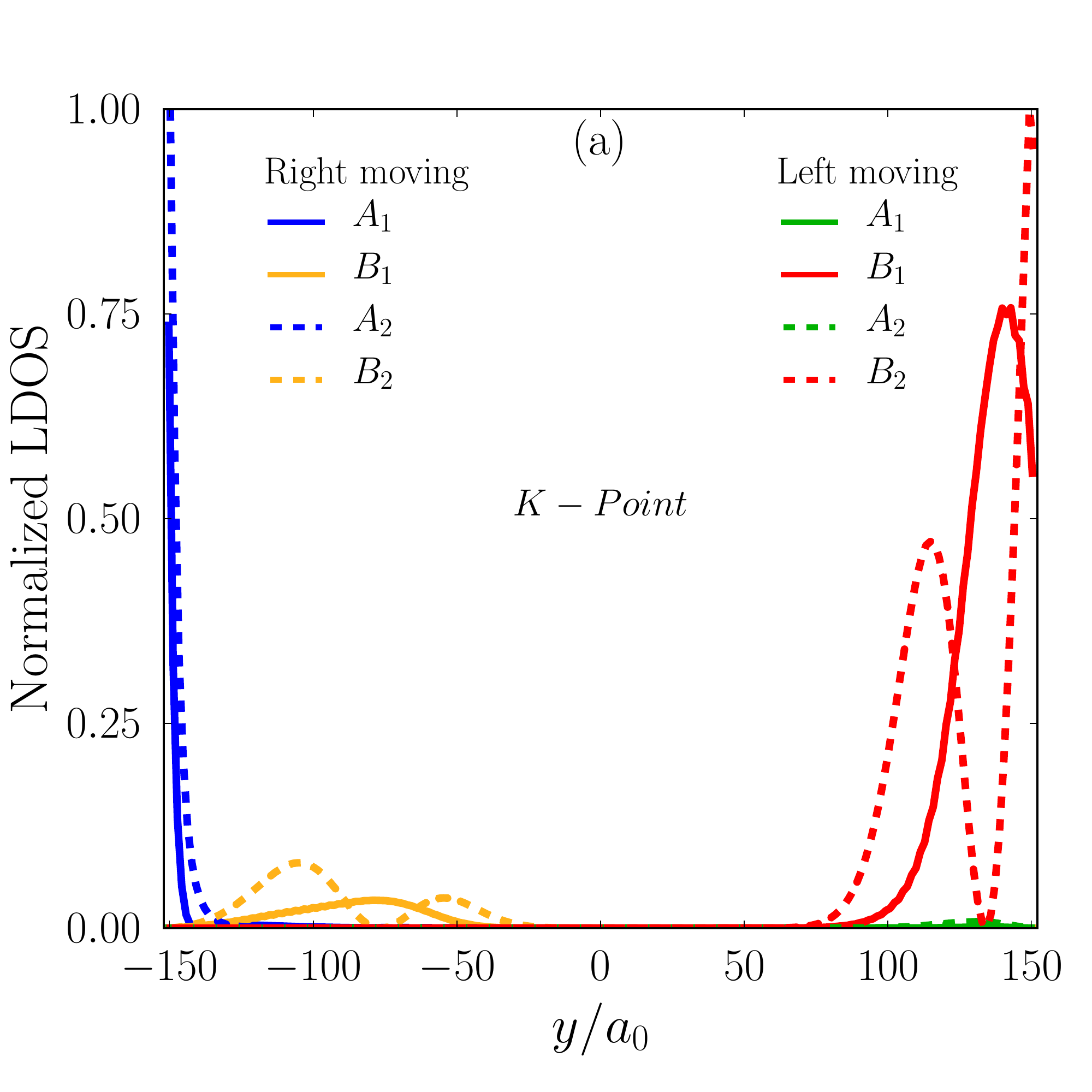}
\includegraphics[width=0.49\linewidth]{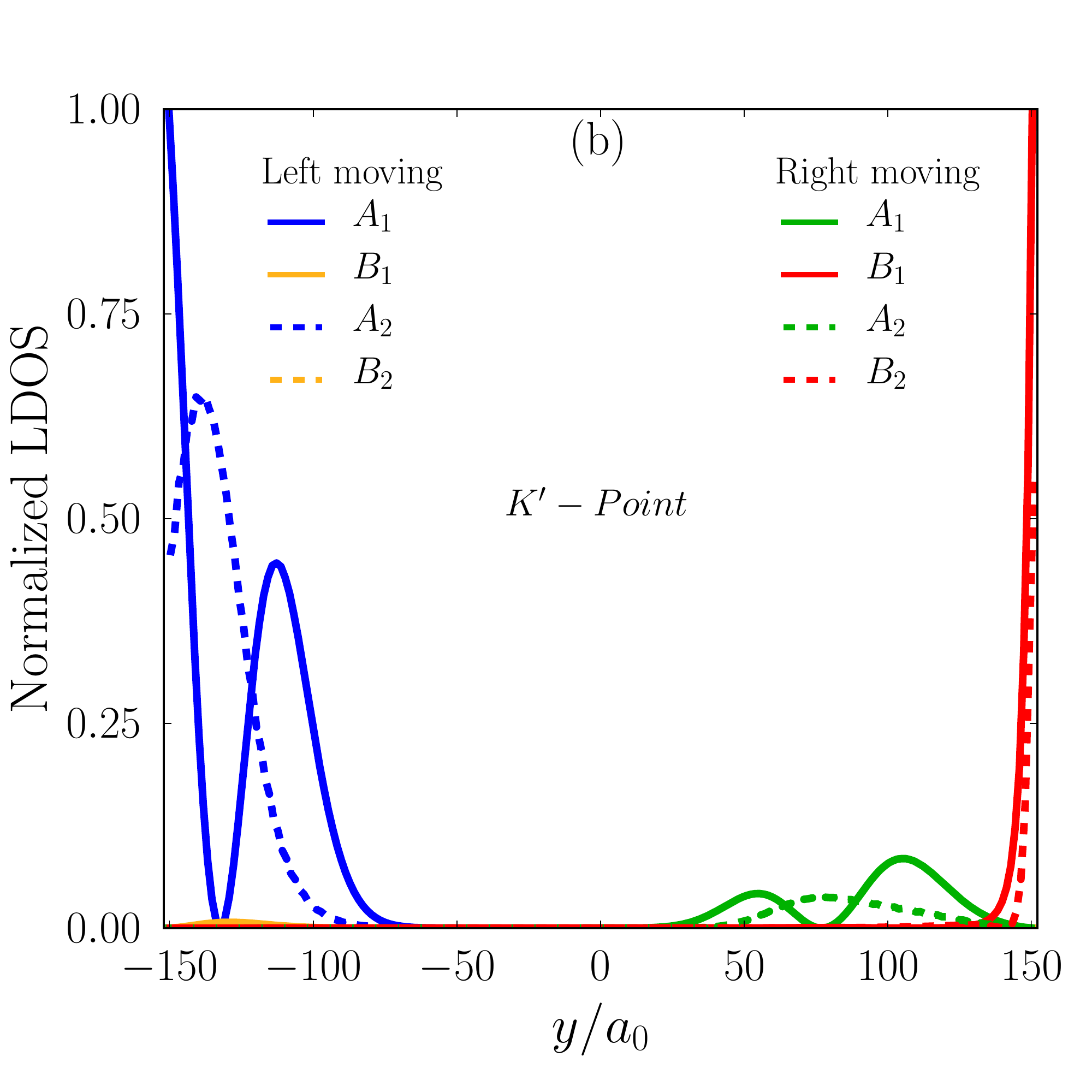}
\includegraphics[width=0.49\linewidth]{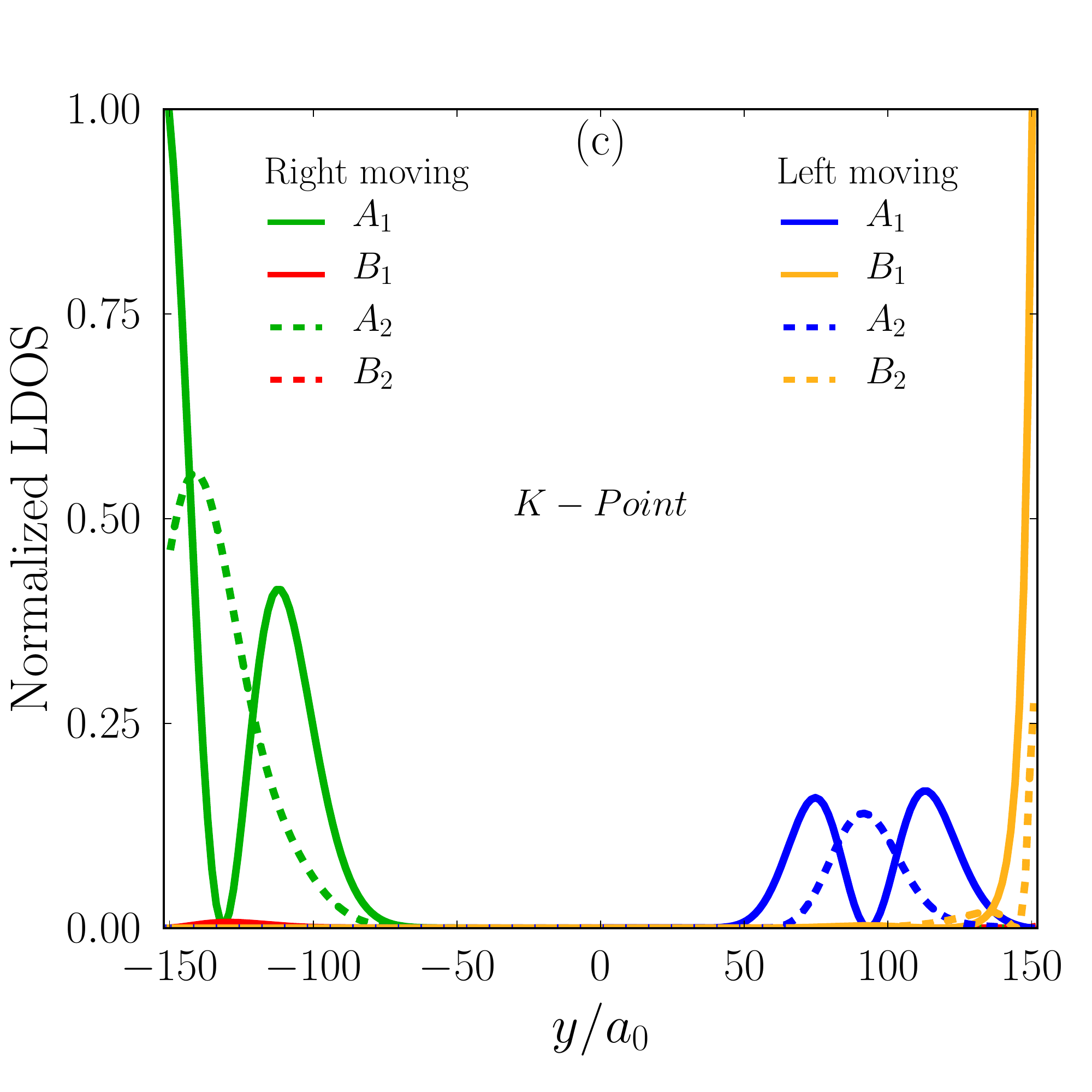}
\includegraphics[width=0.49\linewidth]{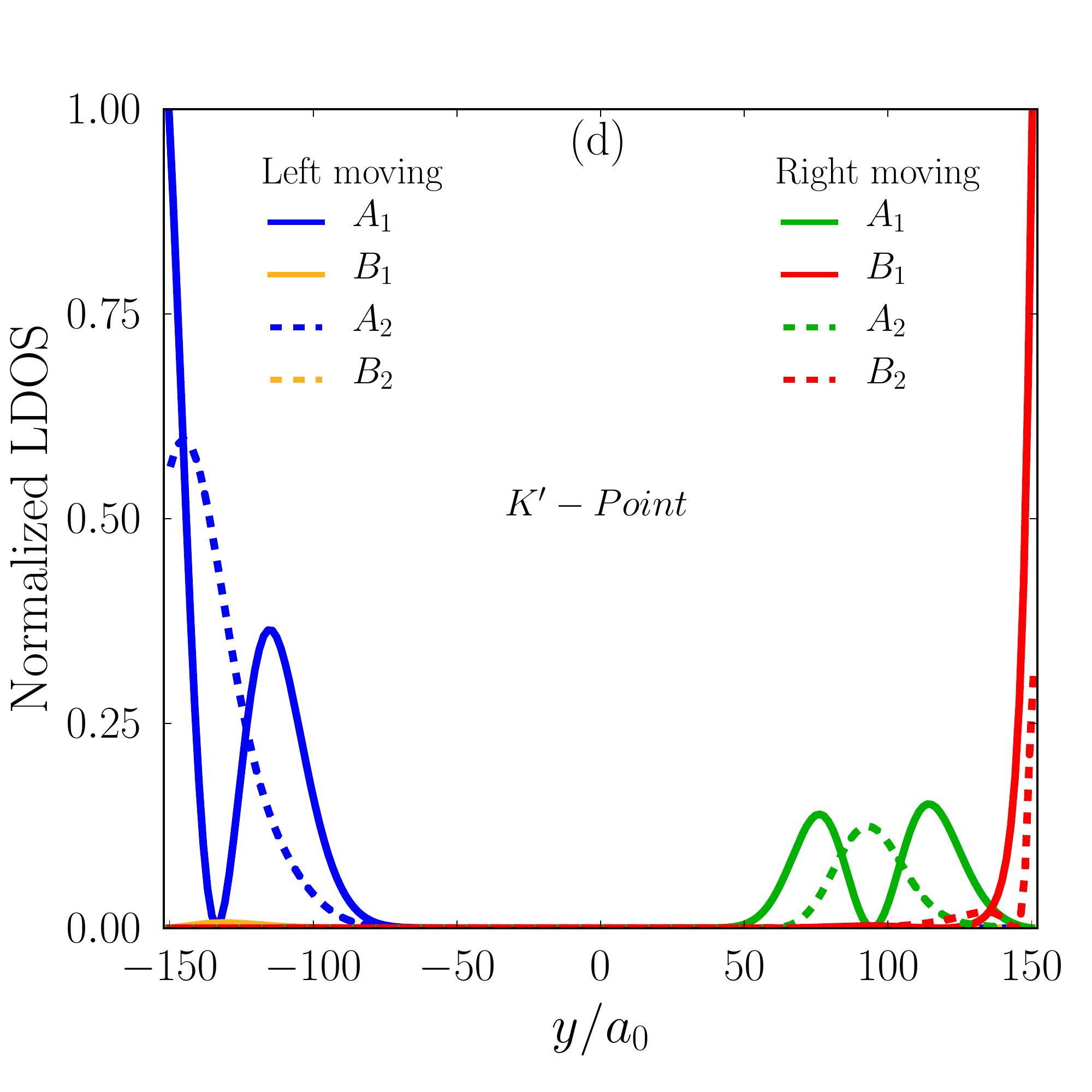}
\caption{(Color online) LLs in a biased BLGNR (a, b) and PLLs in the arc-shaped bending of the biased BLG (c, d).
The edge states in the gap region are localized on different edges and
sublattices for each channel of the edge and electrons in two valleys
circulate in opposite directions. Notice that the right moving refers to ($A_i$, $B_1$) and the left moving denotes to
($B_i$, $A_1$) sublattices at the Dirac points, ${\bm K}$ and ${\bm K'}$ respectively,
$i=2$ for the real magnetic and $i=1$ for the pseudomagnetic fields. In biased BLGNR subjected by the real
magnetic field, a quantum valley Hall effect is expected rather than
the quantum Hall effect when the Fermi energy is in the band gap.
}
\label{fig6}
\end{figure}

The quantized Hall effect is a generic behavior of 2D electron
systems in a strong perpendicular magnetic field. For instance,
biased BLGNR in a real magnetic field has $\sigma_{xy}=0$ plateau
in Hall measurement which originates from the zero charge
transfer~\cite{ref:biased1, ref:biased2} across the edge states.
We calculate the charge circulation direction for each edge and
our numerical results are illustrated in Fig.~\ref{fig6}. Charge
carriers circulate in opposite directions in two valleys which
means that the Hall conductivity of each valley changes in sign
and therefore the total Hall conductivity, $\sigma_{xy}$ vanishes
in a consistence results with experimental
measurements~\cite{ref:biased1} and theoretical
predictions.~\cite{ref:biased2} However, valley Hall conductivity,
which is defined as
${\sigma}^v_{xy}={\sigma}^K_{xy}-{\sigma}^{K'}_{xy}$, is finite
due to the inversion symmetry breaking.~\cite{ref:Xiao07} The
non-zero valley Hall conductivity originates from the opposite
sign of the Berry curvature in two valleys and it is easy to
obtain the Berry curvature as $\Omega_{xy}=-\tau_z 2\Delta
a^4|q|^2/(\Delta^2+a^2 |q|^4)^{3/2}$ where $a=\hbar
v_0^2/\gamma_1$ where $\Delta=u/2$. Consequently, in the biased
BLGNR which is subject to the real magnetic field, a quantum
valley Hall effect is expected rather than the quantum Hall effect
when the Fermi energy is in the band gap. Notice that the state of
the system is either in the quantum Hall (QH) phase or in the
valley polarized quantum Hall (VPQH) phase for higher Landau
levels. The VPQH actually occurs because of the different shifts
of LLs around the two valleys. In strained biased BLG, on the
other hand, pseudomagnetic field gives different features for LDOS
where layer polarization is obtained and the system is in a
quantum valley Hall (QVH) phase when the Fermi energy cuts four
edge modes as it is shown in Fig.~\ref{fig5}. Accordingly, the
VPQH phase is absent in the pseudomagnetic field case which
preserves the time reversal symmetry. For higher energy, there is
a small interval with a trivial insulating gap, without any
accusable bulk and edge states, and then system tends to the QVH
insulator phase due to different sign of the pseudomagnetic field
at two valleys.

\begin{figure}
\includegraphics[width=1\linewidth]{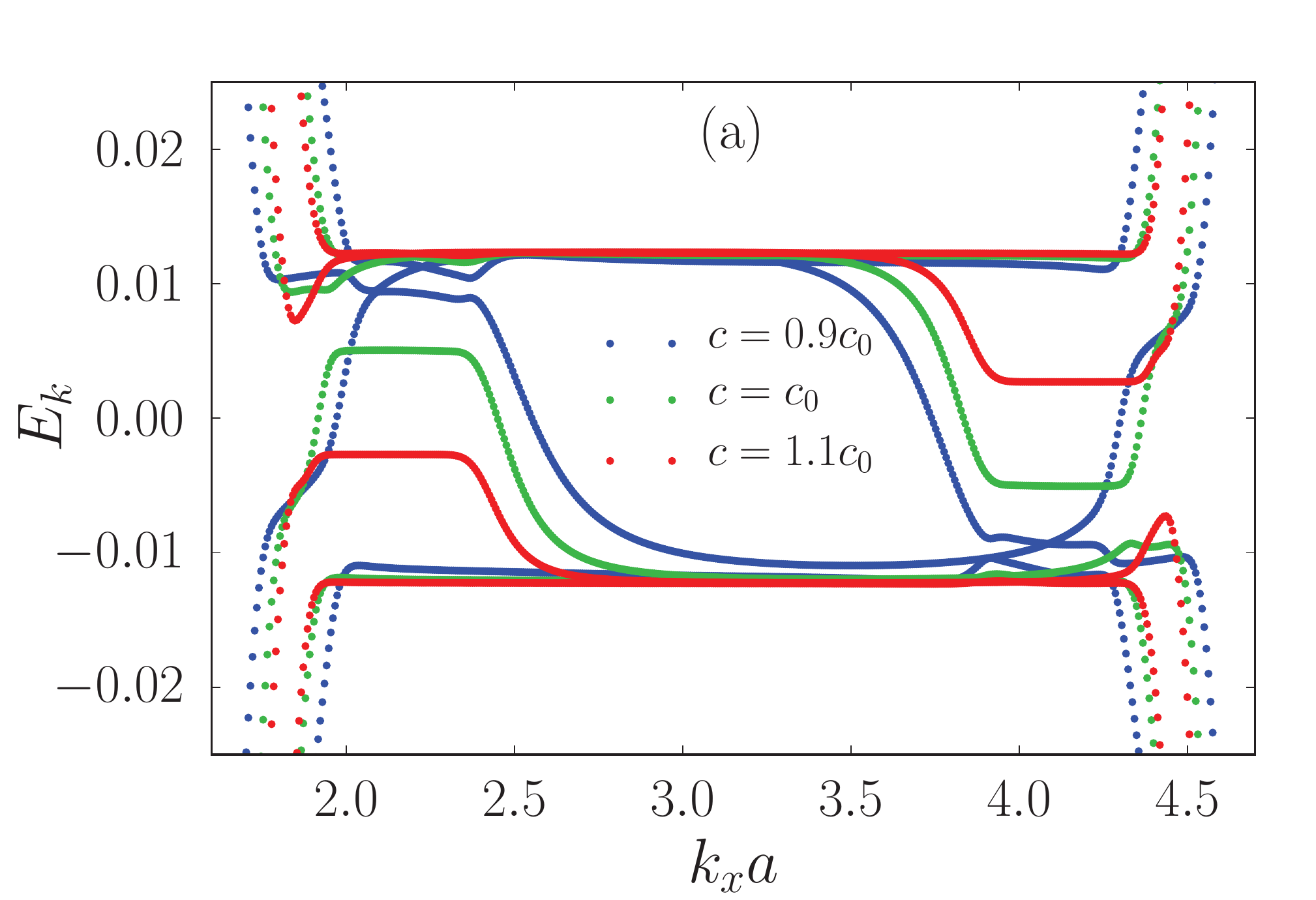}
\includegraphics[width=1\linewidth]{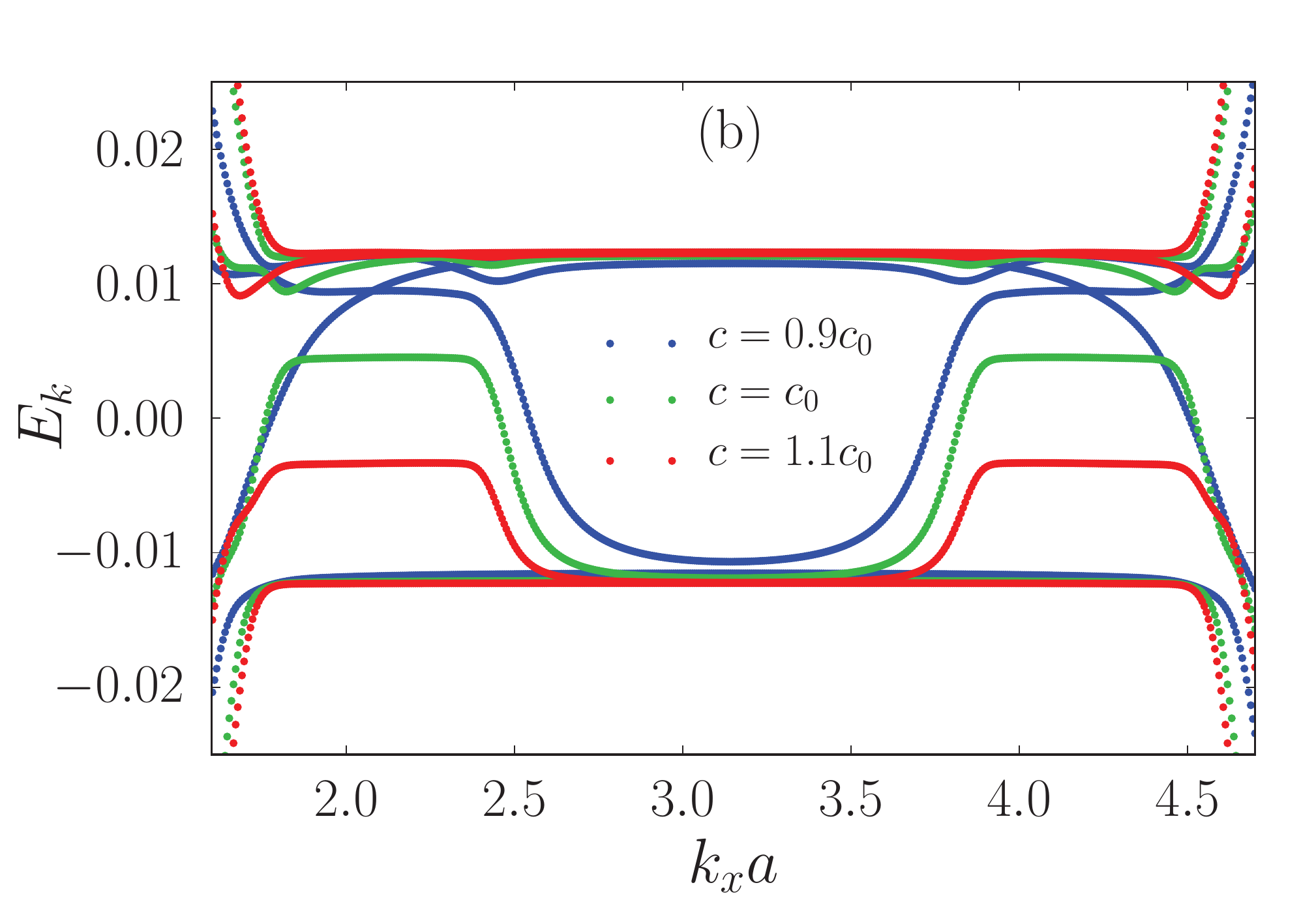}
\caption{(Color online) LLs in a biased BLGNR (a) and PLLs in the arc-shaped bending of the biased BLG (b)
for different values of the interlayer hoping integral for $W=300.5a_0$, $u=\gamma_1/5$, $R=781.3a_0$ and $l_B=23.2a_0$. The energy
dispersion of biased BLGNR profoundly depends on the interlayer
spacing which mainly changes the magnitude of $\gamma_1$. For a
real magnetic field, the large and small values of the interlayer
spacing have two different forms of the edge states for zero Landau
level. There is a crossed edge states inside the band
gap for a larger value of the interlayer, while a
crossing disappears and there is a valley polarization
for the case that the interlayer hopping is
small. There is a topological phase
transition from QVH to VPQH and trivial insulator phase by tuning
the intralayer spacing at a certain value of Fermi energy. The ratio of
$B/{\gamma_1}$ where $B$ is the real magnetic field is the
tuning parameter. In pseudomagnetic
field, on the other hand, the VPQH phase is absent.}
\label{fig7}
\end{figure}

\begin{figure}
\includegraphics[width=1\linewidth]{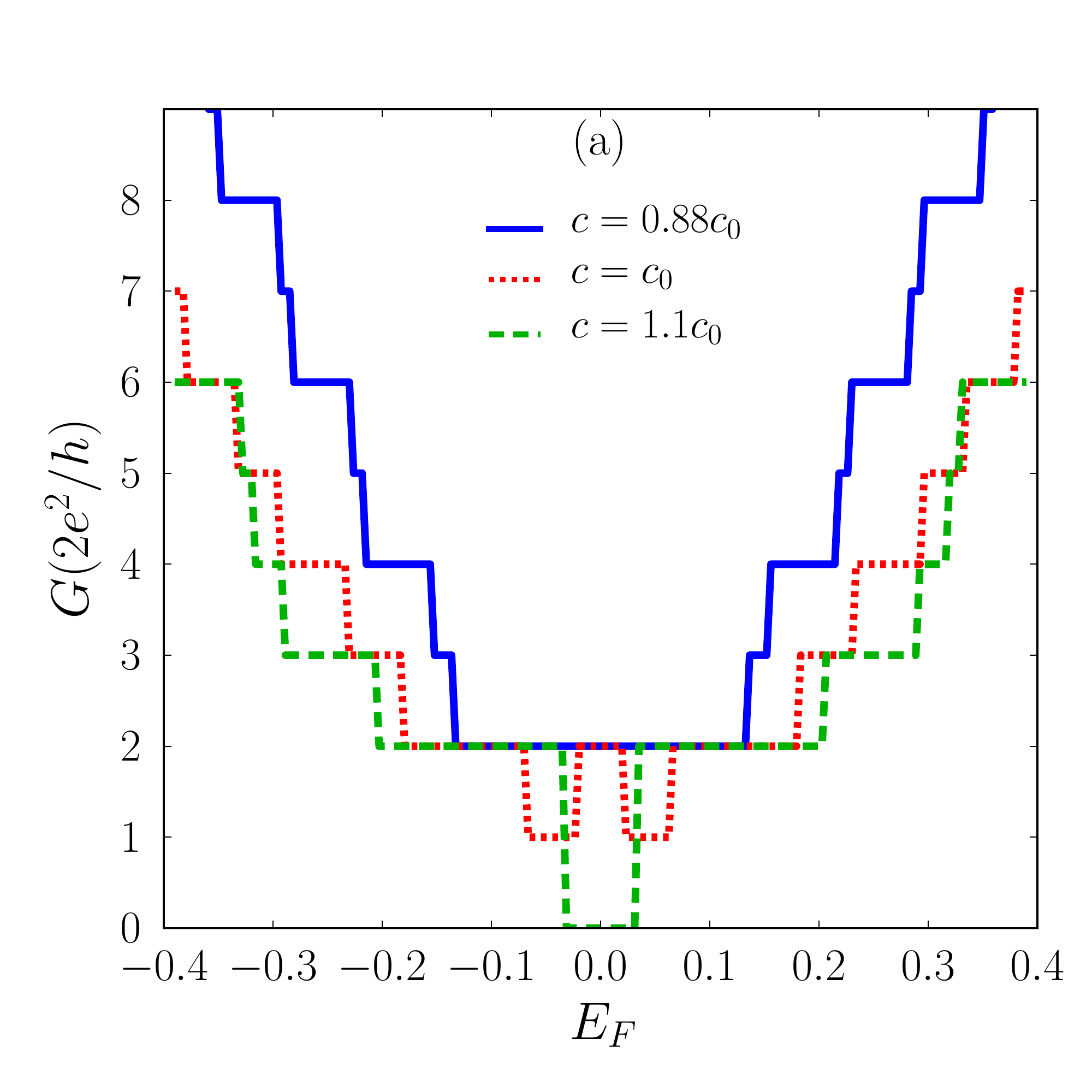}
\includegraphics[width=1\linewidth]{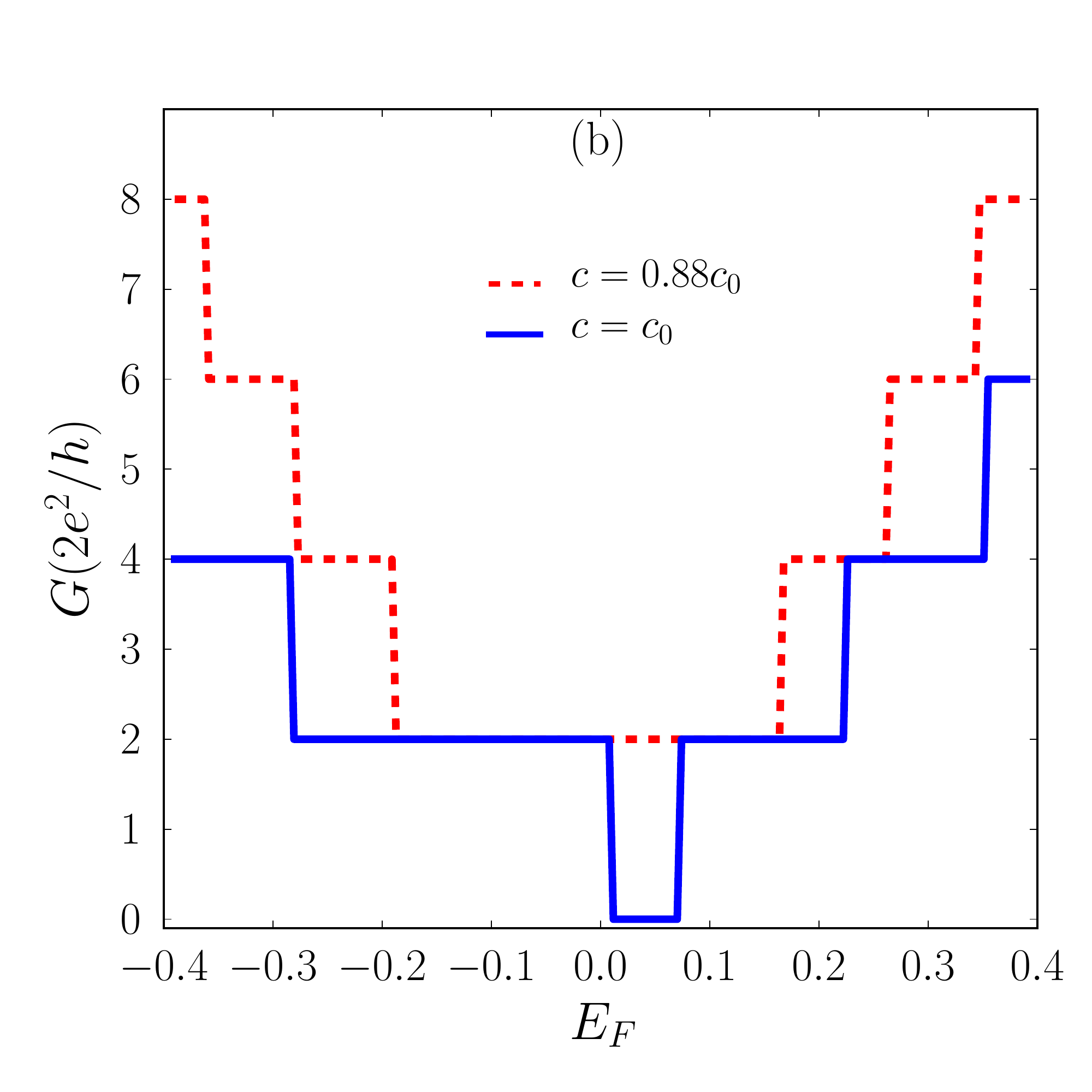}
\caption{(Color online) Conductance for different phases in real (a) and
pseudomagnetic field (b) for $W=150.5a_0$, $R=496.6a_0$, $l_B=21.3a_0$ and $u=\gamma_1/2$.
The energy is measured in units of eV.}
\label{fig8}
\end{figure}

\begin{figure}
\includegraphics[width=1\linewidth]{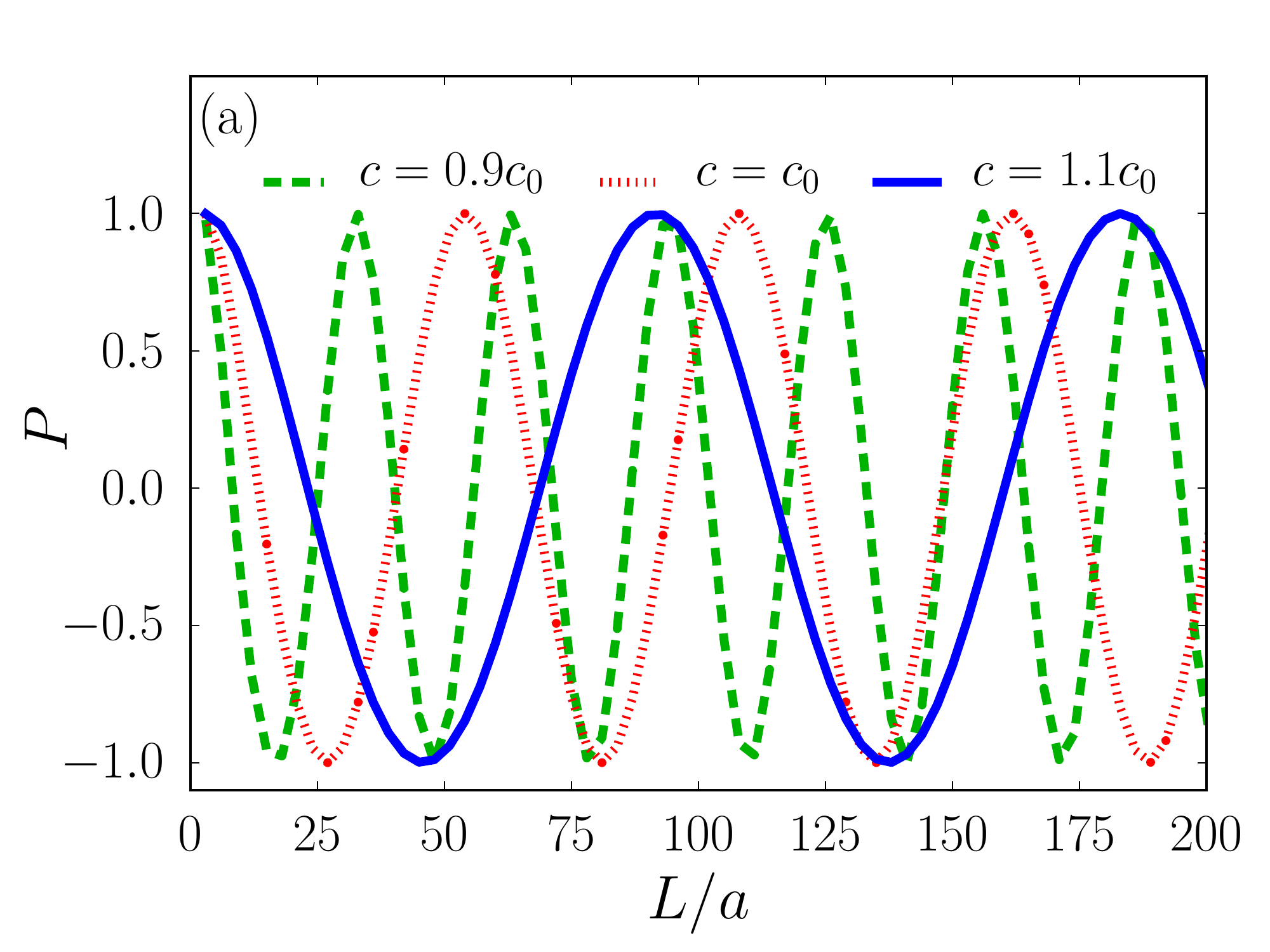}
\includegraphics[width=1\linewidth]{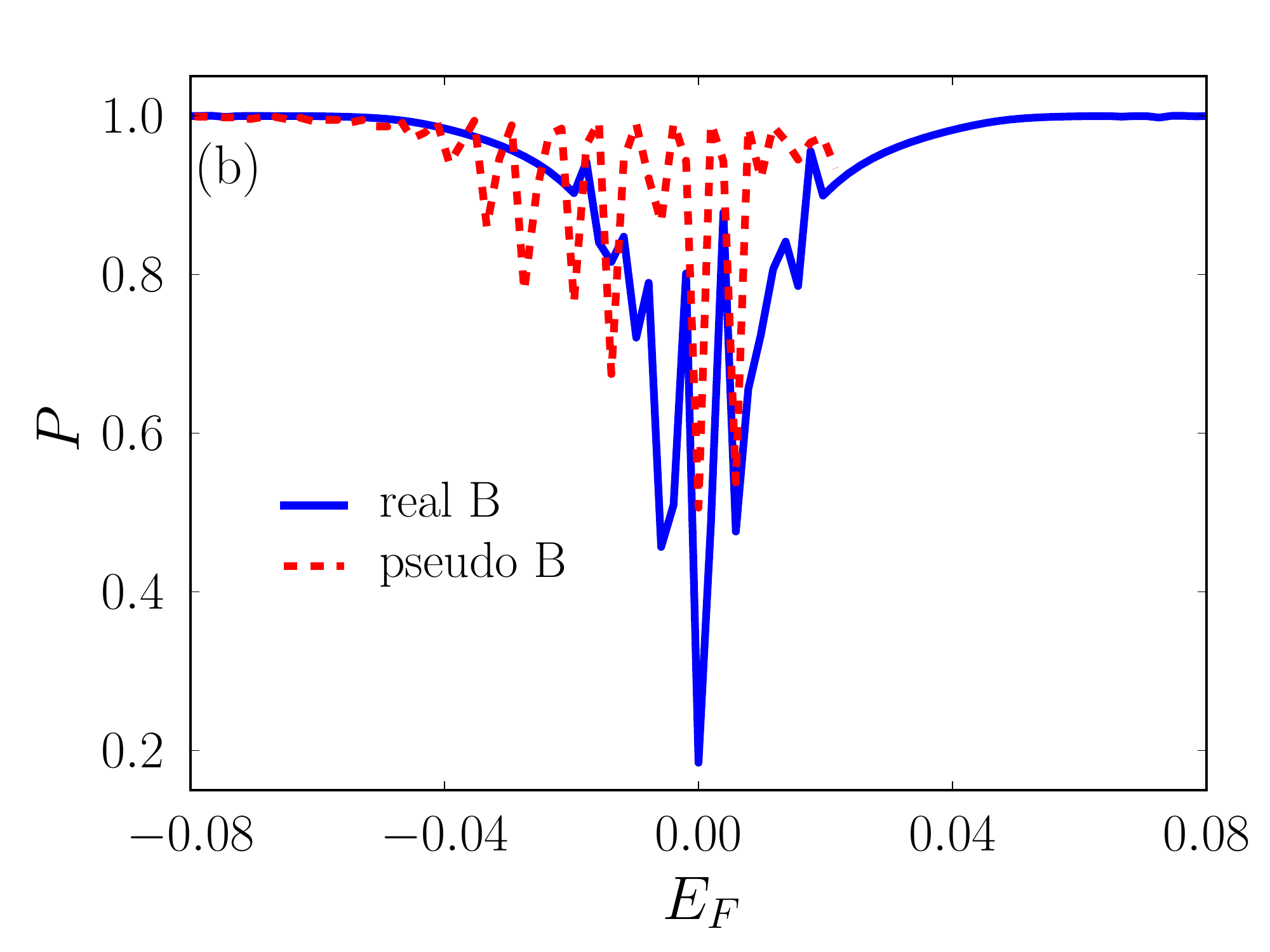}
\caption{(Color online) (a) Layer polarization as a function of channel length for
 different interlayer spacing at real and pseudomagnetic field for
 $W=75.5a_0$, $u=0$, $E=0.01$eV and $a=\sqrt{3}a_0$. The layer-resolved transport shows a periodic behavior of
the polarization without decaying manner. The period of the
oscillation can be tuned by the interlayer spacing and the period
increases with increasing $c$ values. (b) Layer polarization as a function of
the Fermi energy for biased BLGNR in the presence of a real and pseudomagnetic field
for $W=150.5$, $R=451.5a_0$, $l_B=20.3a_0$, $L=200a$, $u=\gamma_1/2$ and $c=c_0$.
The energy is measured in scale of eV.  In the trivial
insulator phase there is no transport channel. In the VPQH phase, on
the other hand, $P\approx 1$ and it indicates layer-conserved
transport. Also the polarization in the QVH phase
in the presence of pseudomagnetic field and the vertical bias is
higher than the value obtained in the QVH in the real magnetic
field. }
\label{fig9}
\end{figure}

In strained graphene the energy band structure is deeply
influenced by the form and the value of hopping correction. We
show in Fig.~\ref{fig7} that the interlayer spacing has a
significant effect on the energy dispersion of a biased BLGNR and
hence we study this effect in the presence of a real and
pseudomagnetic fields. The energy dispersion of the biased BLGNR
depends strongly on the interlayer spacing which mainly changes
the magnitude of $\gamma_1$. For a real magnetic field, the large
and small values of the interlayer spacing have two different
forms of edge states for zero Landau level. There is a crossed
edge states inside the band gap~\cite{ref:biased1} for larger
value of the interlayer, while the crossing disappears and there
is a valley polarization (VP)~\cite{ref:fertig} for the case that
the interlayer hopping is small. Here we demonstrate that there is
a topological phase transition from QVH to VPQH and trivial
insulator phase by tuning the intralayer spacing at a certain value
of the Fermi energy. This feature is shown in Fig.~\ref{fig7}a.
Note that the ratio of $\alpha={B}/{\gamma_1}$ where $B$ is the
real magnetic field is the tuning parameter. The reason for that
is the fact that the zero Landau level depends on the magnetic
field~\cite{ref:fertig}, $E_0(\alpha)$. These results are
interesting because the system reveals three different phases
inside the band gap for different values of the interlayer
spacing. In pseudomagnetic field, on the other hand, the VPQH
phase is absent due to the presence of time reversal symmetry and
corresponding numerical results are displayed in Fig.~\ref{fig7}b.

We calculate the conductance corresponding to the different phases
and our numerical results are shown in Fig.~\ref{fig8}. The
conductance indicates the number of the transport channel in a two
terminal setup. In the presence of the real magnetic field
plateaus are even integer values for QH and QVH phases however
there is an odd integer for VPQH phase due to the breaking of the
valley degeneracy. For the pseudomagnetic field case, the valley
degeneracy preserves in order that plateaus are even integer. For
both the real and pseudomagnetic fields two terminal conductance
is zero in the trivial insulating phase due to the absence of any
modes to carry any kinds of current.

Moreover, in order to investigate the effect of the interlayer
spacing in detail we study the layer-resolved transport properties
of BLGNR by calculating $P$ as a quantity which indicates a
spectrum from fully layer-conserved ($P=1$) to fully layer-flip
($P=-1$) transport. In order to do so, we calculate the
polarization as a function of channel length and thus
Fig.~\ref{fig9}a shows a periodic behavior of the polarization
similar to its spin-analogue~\cite{ref:Pareek02} without showing a
decaying manner. The reason for that is related to the absence of
scattering centers in the ballistic transport. It is obvious that
the period of the oscillation can be tuned by the interlayer
spacing and the period increases with increasing the interlayer
spacing values, $c$. Consequently, strain can tune the
layer-resolved features in BLGNR. Furthermore, we study the
layer-resolved characteristics of BLGNR in the different phases
and the numerical results are shown in Fig.~\ref{fig9}b. Note that
in the trivial insulator phase there is no transport channel so
that the polarization does not have any physical meaning. In the
VPQH phase, on the other hand, $P\approx 1$ and it indicates that
the layer-conserved transport because, as it is shown in
Fig.~\ref{fig6}, the right moving transport channel for zero
Landau level edge states are mainly localized on different layers
for the two valleys. While, in the QVH phase, resulting from the
real magnetic field and vertical gate voltage, the polarization
reduces although the transport is mainly due to the layer-conserve
case. Also the polarization in the QVH phase in the presence of
the pseudomagnetic field and the vertical bias is higher than the
value in the QVH for the real magnetic field case. This feature
can be understood by looking at the LDOS presented in
Fig.~\ref{fig6} in which the right moving edge states are mostly
localized in the bottom layer for the zero pseudo
Landau level. There is the trivial insulator phase inside the gap
for $E_{\rm F}>0.02eV$ and the pseudomagnetic case for which
transport channels are absented. Indeed, in our special lead
structures, the resonating feature of the polarization and
conductance comes from the hopping mismatch between the leads and
the device region too. This mismatch is a source of a quantum
interference effect such as Fabry–P\'{e}rot
resonances.~\cite{ref:Nazarov}

\section{conclusion}

We have analyzed in detail the electronic properties of bilayer
graphene by putting together the effects of deformations, real
magnetic field and perpendicular gate voltage by modifying a
lattice model Hamiltonian. We have proposed a lattice model
Hamiltonian to explore bilayer graphene nanoribbon under the
combined effect of deformations, real magnetic fields and gate
voltages. The lattice model Hamiltonian, which we have proposed,
provides a quite good description of the edge states and we have
shown that the pseudo Landau levels in the bulk are no longer
dispersive. We have considered a strained bilayer graphene
nanoribbon by deforming it in the arc-shaped bending structure in
the presence of a real magnetic field and found that the
zero-energy is chiral. Moreover, the effective total fields acting
on electrons from the two-valleys are different which results in a
valley polarization.

We have studied the edge states inside the band gap in the
presence of a real magnetic field and explored a valley polarized
quantum Hall effect, which occurs due to the different shifts of
Landau levels around the two valleys, in biased bilayer graphene
nanoribbon systems. We have also demonstrated that, in biased
bilayer graphene which is subjected to a real magnetic field, the
energy dispersion depends on the interlayer spacing and a
topological phase transition from quantum valley Hall to valley
polarized quantum Hall can occur by tuning the interlayer spacing
between two layers. The valley polarized quantum Hall goes toward to a
trivial insulator phase with decreasing of the interlayer spacing. In
strained biased bilayer graphene, on the other hand, the system is
in a quantum valley Hall phase.

The numerical calculations that we have employed to carry out the
conductance is the recursive Green's function. Based on that, we
have investigated the effect of the interlayer spacing on the
layer-resolved transport in bilayer graphene by using layer
polarization and its value confirms our obtained phases. We show
that the layer polarization in the presence of a pseudomagnetic
field is larger than that expected for the value obtained in
the presence of the real magnetic filed. We have found that the
plateaus of the conductance are even integer values for quantum
Hall and quantum valley Hall phases however there is an odd
integer for valley polarized quantum Hall phase due to the
breaking of the valley degeneracy in the presence of the real
magnetic field. For the pseudomagnetic field case, the valley
degeneracy preserves in order that plateaus are even integer. It
would be interesting to investigate all mentioned phases in the
presence of the electron-electron interactions. Valley
polarization quantum Hall phase and layer polarization physics, in
 such way that we proposed, can have very important implications
on the electronic properties of bilayer graphene ribbons and other
nanostructures.

\begin{acknowledgments}

R. A. would like to thank the international center for theoretical
physics, ICTP, for its hospitality during the period when the last
part of this work was carried out. We thank A. Naji for his useful
comments.
\end{acknowledgments}

\end{document}